\documentclass{article}
\pdfoutput=1
\usepackage{jheppub}
\notoc

\usepackage{graphicx}
\usepackage{hyperref}
\usepackage[sort&compress]{natbib} 
\usepackage[english]{babel}
\usepackage{amsmath}
\usepackage{amssymb}

\newcommand{\de}{\delta} 

\newcommand{\eref}[1]{Eq.~(\ref{#1})}
\newcommand{\fref}[1]{Fig.~\ref{#1}}
\newcommand{\nnnl}{\nonumber\\}	

\newcommand{\beq}{\begin{eqnarray}}
\newcommand{\eeq}{\end{eqnarray}}

\newcommand{\LL}{{\Lambda^2}}
\newcommand{\LQ}{{\Lambda_\mathrm{QCD}^2}}

\begin{document}
\title{Spurious divergences in Dyson-Schwinger equations}
\date{\today}

\author[a,b]{Markus Q. Huber,}

\author[a,b]{Lorenz von Smekal}

\affiliation[a]{Institut f\"ur Theoretische Physik, Justus-Liebig-Universit\"at, 35392 Gie\ss en, Germany}
\affiliation[b]{Institut f\"ur Kernphysik, Technische Universit\"at Darmstadt, 64289 Darmstadt, Germany}



\emailAdd{markus.huber@physik.tu-darmstadt.de}
\emailAdd{lorenz.smekal@physik.tu-darmstadt.de}

\abstract{We revisit the treatment of spurious ultraviolet divergences 
in the equation of motion of the gluon propagator caused by a momentum
cutoff and the resulting violation of gauge invariance. With present 
continuum studies of the gluon propagator from its Dyson-Schwinger
equation reaching the level of quantitatively accurate descriptions, it
becomes increasingly important to understand how to subtract these
spurious divergences in an unambiguous way. Here we propose such a
method. It is based entirely on the asymptotic perturbative behavior
of the QCD Green's functions without affecting non-perturbative
aspects such as mass terms or the asymptotic infrared behavior.   
As a particular example, this allows us to assess the possible 
influence of the tadpole diagram beyond perturbation theory.
Finally, we test this method numerically by solving the system of
Dyson-Schwinger equations of the gluon and ghost propagators. 
}

\maketitle

\section{Introduction}

The basic entities of a local quantum field
theory are its Green's functions. They can be calculated with various
methods ranging from perturbation theory to Monte-Carlo simulations on
discretized space-time lattices or functional equations, e.g., via the
functional renormalization group or their  equations of motion, the
Dyson-Schwinger equations (DSEs). Studies of gluonic Green's
functions in quantum chromodynamics (QCD) using their DSEs have a long
history. For a selection of such studies see
\cite{vonSmekal:1997is,vonSmekal:1997vx,Atkinson:1997tu,Alkofer:2000wg,Fischer:2002eq,Fischer:2002hn,Fischer:2003rp,Fischer:2006ub,Alkofer:2008tt,Boucaud:2008ji,Aguilar:2008xm,Fischer:2008uz,Boucaud:2010gr,Pennington:2011xs,Maas:2011se,Binosi:2012sj,Strauss:2012dg,Huber:2012kd,Fischer:2012vc,Muller:2013pya,Hopfer:2013np,Muller:2013tya,Fischer:2013eca,Aguilar:2013xqa,Blum:2014gna,Aguilar:2014tka,Eichmann:2014xya}.
Over the past two decades there has been a steady development
that pushed the technical limits and thereby improved the quantitative
reliability of the results. Many conceptual problems were solved, but
a perpetual challenge for many DSE practitioners remained in how to
deal with spurious ultraviolet (UV) divergences in a practically
feasible yet unambiguous way. Such divergences occur, when a regularization is used 
that violates gauge invariance. The mostly for its technical
simplicity used O(4) invariant Euclidean UV cutoff is such a
regularization. Feasible alternatives without this problem are
scarce. Dimensional regularization, for example, which does not
have this problem and which is therefore best suited for analytic
calculations (see, e.g.\ \cite{Pascual:1984}), is not easily
implemented in numerical computations
\cite{Gusynin:1998se,Phillips:1999bf}. Thus, from a practical 
point of view, it is still desirable to cure this artificial
problem of spurious divergences with a simple UV cutoff in an
unambiguous way.  In contrast to the unavoidable logarithmic UV
divergences of a renormalizable theory in four space-time dimensions,
they cannot be removed by standard multiplicative renormalization. 

Several ways to remove these spurious divergences were proposed in the
literature and used in the past. We review some of them in
Sec.~\ref{sec:quadDivs}. However, depending on which method is used,
the results can vary to some extent in the non-perturbative
regime. At the moment, these variations may not be dramatic but still
within the overall uncertainties of present approximations and the
systematic truncation errors, especially in the gluon propagator DSE
in which the explicit two-loop contributions are typically neglected. 
As the systematics in the truncations and hence the quantitative reliability
of DSE results are steadily improving, however, we will inevitably
reach a point at which the variations due to the different subtractions of  
spurious UV divergences will matter as well. This motivates to search
for a better understanding of how to subtract them without affecting
non-perturbative aspects such as dynamically generated mass terms, or
condensate contributions, and the infrared behavior.  

Often the ambiguity in the non-perturbative regime is directly
reflected in the appearance of a new parameter. At first sight this might
also seem to be the case for the method proposed here. However, upon closer
inspection, in particular of the asymptotic perturbative behavior of
the propagators in the ultraviolet, where calculations can be done
analytically, the subtraction of the spurious divergences in
the gluon propagator can be fixed unambiguously by requiring that it
must not introduce a mass term. We
explicitly demonstrate this and verify that the subtracted spurious
contributions remain the same in the fully non-perturbative
calculation. 
Our perturbative subtraction method moreover allows to assess possible 
contributions from the tadpole diagram in the gluon DSE, which are
usually discarded because they vanish in perturbation
theory, but which can contribute beyond that.

Before we illustrate in Sec.~\ref{sec:quadDivs} how the spurious
(quadratic) divergences emerge, we give a short review of the gluon
propagator DSE in Sec.~\ref{sec:glProp}. Some methods used in the
literature to subtract these divergences are listed in
Sec.~\ref{sec:quadDivs}. In Sec.~\ref{sec:sub_procedure} we detail our
new method which is used to obtain the results presented in
Sec.~\ref{sec:effects_gluon}. We summarize our work in
Sec.~\ref{sec:summary_conclusions}. Two appendices contain further
details on the derivation of the subtraction term.

\section{The gluon propagator DSE}
\label{sec:glProp}

\begin{figure}[tb]
 \includegraphics[width=\textwidth]{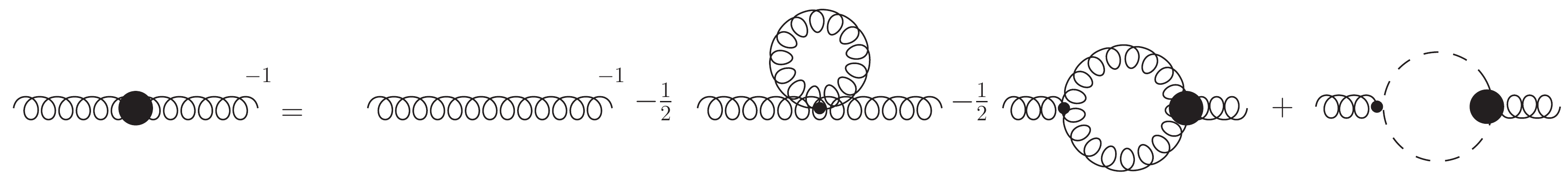}
 \caption{\label{fig:gl_DSE}Standard truncation of the gluon
   propagator DSE to explicit one-loop structures which contains the bare
   inverse gluon propagator, the gluon and ghost
   loops, and the tadpole (second diagram on the right), which is normally also dismissed. Curly (dashed)
   lines are gluon (ghost) propagators. Thick (thin) filled circles
   represent dressed (bare) vertices. All internal propagators are 
   dressed as well.}  
\end{figure}

The truncated DSE of the gluon propagator is shown diagrammatically in
\fref{fig:gl_DSE}. We consider the widely used truncation in which all
explicit two-loop diagrams are neglected. Unlike most previous studies
we have maintained also the tadpole contribution here for now, however, which
we will discuss in more detail below.
The gluon and ghost propagators in the Landau gauge 
are given by (color indices suppressed)
\begin{align}
 D_{gl,\mu\nu}(p^2):=D(p^2) P_{\mu\nu}(p^2):=P_{\mu\nu}(p)\frac{Z(p^2)}{p^2}, \quad D_{gh}(p^2):=-\frac{G(p^2)}{p^2},
\end{align}
where $P_{\mu\nu}$ is the transverse projector and $G(p^2)$ and
$Z(p^2)$ are the ghost and gluon dressing functions. It is by now well
understood \cite{Fischer:2008uz} that their DSEs admit a family of
solutions which are classified by the value of the inverse ghost
dressing function in the infrared, $G^{-1}(0)$. Finite positive 
values give rise to the so-called {\em decoupling} solutions \cite{Dudal:2007cw,Aguilar:2008xm,Boucaud:2008ji,Fischer:2008uz,Alkofer:2008jy} with 
\begin{equation} 
 Z(p^2) \sim p^2/M^2 \, , \;\; G(p^2) \to \mbox{const.}, \;\;
 \mbox{for} \;\; p^2 \to 0\, ,
\end{equation}
and, as a limiting case with $G^{-1}(0) =0$, one obtains the {\em
  scaling} solution \cite{vonSmekal:1997is,Pawlowski:2003hq},  
\begin{equation} 
 Z(p^2) \sim (p^2)^{2\kappa} \, , \;\; G(p^2) \sim (p^2)^{-\kappa} , \;\;
 \mbox{for} \;\; p^2 \to 0\, ,
\end{equation}
with a positive exponent $\kappa < 1$ which can be calculated
analytically under a certain regularity assumption on the ghost-gluon
vertex \cite{Lerche:2002ep,Huber:2012zj}. It is given by 
$\kappa = (93 - \sqrt{1201})/98 \approx 0.6$ \cite{Lerche:2002ep,Zwanziger:2001kw}. Only decoupling is seen
on the lattice, e.g., \cite{Cucchieri:2007md,Cucchieri:2008fc,Sternbeck:2007ug,Bogolubsky:2009dc,Oliveira:2012eh,Sternbeck:2012mf}, and the reasons for that are fairly well understood
\cite{vonSmekal:2008ws}. 

An equation for the gluon dressing function $Z(p^2)$ is obtained by
a suitable projection of its DSE.
Mainly for bookkeeping purposes we use a generalized projector as in Refs.~\cite{Fischer:2002eq,Fischer:2002hn}:
\begin{align}
 P^{\zeta}_{\mu\nu}(p)=g_{\mu\nu}-\zeta\frac{p_\mu p_\nu}{p^2}. \label{eq:Pzeta}
\end{align}
In a manifestly transverse truncation to the full gluon DSE the
solution will not depend on the parameter $\zeta $ introduced here.
Conversely, the required $\zeta$-independence can be used as a valuable test
of specific truncations in numerical studies, see the discussion in
Ref.~\cite{Fischer:2010is}.  
Here we will simply set $\zeta=1$ for the transverse projector later on.
For now, however, we contract the gluon DSE with the projector in \eref{eq:Pzeta}:
\begin{align}
 \frac{1}{Z(p^2)}&=\widetilde{Z}_3+N_c\,g^2\,Z_4\int_q Z(q^2)K_{Z}^{tad,\zeta}(p,q)\nnnl
 &+N_c\,g^2\,\widetilde{Z}_1\int_q G(q^2)G((p+q)^2) K_{Z}^{gh,\zeta}(p,q)\nnnl
 &+N_c\,g^2\,Z_1\int_q Z(q^2)Z((p+q)^2) K_{Z}^{gl,\zeta}(p,q)D^{A^3}(p^2,q^2,(p+q)^2).\label{eq:Z}
\end{align}
The integral measure in \eref{eq:Z} is defined as $\int_q=\int
d^4q/(2\pi)^4$ and the kernels  (with $x=p^2$, $y=q^2$ and
$z=(p+q)^2$) are given by
\begin{align}
 K_{Z}^{gh,\zeta}(p,q)&=\frac{x^2 
\left(\zeta - 2 - 4\, \eta\hat\eta \, (\zeta-1)  \right)
+2 x (y+z)-\zeta  (y-z)^2}{12 x^2 y z}, \label{eq:KZgh}\\
 K_{Z}^{gl,\zeta}(p,q)&=\frac{z^2 \zeta }{24 x^2 y^2}+\frac{z (5 x-x \zeta +4 y \zeta )}{12 x^2 y^2}+\frac{x^2 (-19+\zeta )+2 x y (-17+\zeta )-18 y^2 \zeta }{24 x^2 y^2}\nnnl
 &+\frac{(x-y)^2 \left(x^2+10 x y+y^2 \zeta \right)}{24 x^2 y^2 z^2}+\frac{4 x^3+x y^2 (-17+\zeta )+4 y^3 \zeta -x^2 y (15+\zeta )}{12 x^2 y^2 z},\\
 K_{Z}^{tad,\zeta}(p,q)&=-\frac{\zeta  x^2-2 x ((18-5 \zeta ) y+\zeta  z)+\zeta  (y-z)^2}{12 x^2 y^2}.
\end{align}
where we used the tree-level ghost-gluon vertex of the general
covariant gauges of
Refs.~\cite{Curci:1976ar,Baulieu:1981sb,ThierryMieg:1985yv,Alkofer:2003jr},  
\begin{align}
 \Gamma^{A\bar{c}c,abc}_\mu(k;p,q)=i\,g\,f^{abc} \left( \eta \,p_\mu - \hat{\eta}\, q_\mu\right).
\end{align}
It involves a second gauge parameter $\eta$ with $\hat\eta \equiv 1 - \eta$
such that standard Faddeev-Popov theory corresponds to $\eta=1$,
$\hat\eta=0$, its mirror image after Faddeev-Popov conjugation to
$\eta=0$, $\hat\eta=1$, and the ghost-antighost symmetric
Curci-Ferrari gauges to $\eta=\hat\eta=1/2$
\cite{vonSmekal:2008en}. In Landau gauge there is 
no such distinction, however, and the gluon propagator must be
independent of $\eta$ as well \cite{Lerche:2002ep}.

The renormalization constants of gluon and ghost propagator are denoted
by $Z_3$ and $\widetilde{Z}_3$ and those for the ghost-gluon,
three-gluon and four-gluon vertices by $\widetilde{Z}_1$, $Z_1$ and
$Z_4$. In the following we will use the \textit{MiniMOM} scheme
\cite{vonSmekal:2009ae} which is defined such that 
$\widetilde{Z}_1=1$ in Landau gauge as in minimal subtraction
schemes, and which is also valid for all $\eta$.\footnote{At one-loop
  level with minimal subtraction this was already observed in
  \cite{Baulieu:1981sb}. In general it follows from a Slavnov-Taylor
  identity as can be shown along the lines of Ref.~\cite{Wschebor:2007vh}.} 
The truncated gluon propagator DSE contains two dressed three-point
functions. The dressed ghost-gluon vertex has already been set to its
tree-level counterpart in the expressions above. 
As
an approximation this is well justified by the overall
comparatively small deviations of the fully momentum dependent vertex
from its tree-level form   
\cite{Ilgenfritz:2006he,Cucchieri:2008qm,Schleifenbaum:2004id,Boucaud:2011eh,Fister:2011uw,Huber:2012kd,Aguilar:2013xqa,Pelaez:2013cpa} 
which induce only minor changes in the propagators likewise
\cite{Huber:2012kd,Aguilar:2013xqa}. We can furthermore see in \eref{eq:KZgh}
explicitly that the relevant transverse part of the gluon propagator
obtained for $\zeta= 1$ remains $\eta$-independent in this truncation
\cite{Lerche:2002ep}.    

For the dressed three-gluon vertex we
have furthermore assumed that one can restrict its form in a first
approximation to that of the tree-level vertex which was shown to provide the
dominant tensor structure for $\zeta=1$ \cite{Eichmann:2014xya}.
Consequently, we use an ansatz, 
\begin{align}
 \Gamma^{A^3,abc}_{\mu\nu\rho}(p,q,k)=i\,g\,f^{abc}D^{A^3}(p^2,q^2,k^2)\left((q-p)_\rho g_{\mu\nu} + perm. \right).
\end{align}
The form of the model dressing function $D^{A^3}(p^2,q^2,k^2)$ will be specified
when we need it below.

\section{Spurious divergences in the gluon propagator DSE}
\label{sec:quadDivs}

The existence and type of spurious divergences in the gluon propagator
DSE can be inferred from analyzing the UV behavior of \eref{eq:Z}
\cite{Brown:1988bm,vonSmekal:1997vx,Atkinson:1997tu}. This is done by
replacing all dressing functions in the loop diagrams by their
leading perturbative expressions \cite{vonSmekal:1997vx,Atkinson:1997tu}:
\begin{align}
G_{UV}(x)&=G(s)\left(1+\omega \ln\left(\frac{x}{s}\right)\right)^\delta=G(s)\left(\omega \, t_x \right)^\delta,\label{eq:G-UV}\\
 Z_{UV}(x)&=Z(s)\left(1+\omega \ln\left(\frac{x}{s}\right)\right)^\gamma=Z(s)\left(\omega \, t_x \right)^\gamma,\label{eq:Z-UV}
\end{align}
where $t_x=\ln(x/\LQ)$ and $\Lambda_\mathrm{QCD}^2=s\,e^{-1/\omega}$;
$\delta=-9/44$ and $\gamma=-13/22$ are the propagator anomalous
dimensions at one-loop level,
$\omega=11\,N_c\,\alpha(s)/12/\pi=\beta_0 g^2(s)$ with 
$\alpha(s)=g^2(s)/(4\pi)$ denoting the
strong coupling at some sufficiently large reference scale $s$ in the
UV, $N_c$ is the number of colors and $\beta_0=11N_c/3/(4\pi)^2$ is
the one-loop coefficient of the $\beta$-function. As in
Refs.~\cite{Huber:2012kd} we use the following
expression for the three-gluon vertex in the UV analysis,
\begin{align}
 D^{A^3}_{UV}(p^2,q^2,k^2)&= G_{UV}(\overline{p}^2)^\alpha Z_{UV}(\overline{p}^2)^\beta,\label{eq:DAAA-UV}
\end{align}
with $\overline{p}^2 = (p^2+q^2+k^2)/2$. The exponents $\alpha$ and
$\beta$ are constrained by $\alpha \delta+\beta
\gamma=\delta-\gamma=17/44$ so that they reproduce the correct
anomalous dimension of the vertex. In our numerical calculations we
choose $\alpha=-17/9$ and $\beta=0$, with which \eref{eq:DAAA-UV}
tends to a constant in the infrared (IR) for decoupling
solutions~\cite{Huber:2012kd}. As usual, we furthermore replace the
three-gluon vertex renormalization constant $Z_1$ by a momentum
dependent function $D^{A^3}_{RG}(p^2,q^2,k^2)$ in order to restore the
correct one-loop running of the gluon dressing function resulting from
the truncated DSE \cite{vonSmekal:1997vx,Huber:2012kd}. For the UV
analysis it suffices to set $D^{A^3}_{RG}(p^2,q^2,k^2)=D^{A^3}_{UV}(p^2,q^2,k^2)$. 

These expressions are inserted into \eref{eq:Z}. To extract the
leading contributions to the right hand side of \eref{eq:Z} from large 
 loop momenta we furthermore assume $q\gg p$ which allows us to replace $G(z)$ and $Z(z)$ by $G(y)$ and $Z(y)$. Introducing a UV
 cutoff $\Lambda$ and performing the angle integrations then yields
\begin{align}
 \frac{1}{Z_{UV}(p^2)}&= Z_3 - Z_4\frac{N_c\,g^2}{64\pi^2}\frac{1}{x}\,\int_x^{\Lambda^2} dy \left(3 (\zeta -4)\right) Z_{UV}(y)\nnnl
& +\frac{N_c\,g^2}{192\pi^2}\,\int_x^{\Lambda^2} dy \frac{x
   \left(\zeta-2-4 \, \eta \hat\eta \, (\zeta-1)\right)-(\zeta -4) y}{x y} G_{UV}(y)^2\nnnl
& +\frac{N_c g^2}{384 \pi^2}\int_x^{\Lambda^2} dy \frac{7 x^2+12  (-4+\zeta ) y^2-2 x y (24+\zeta ) }{ x y^2}G_{UV}(y)^{2\alpha} Z_{UV}(y)^{2+2\beta}+\ldots,\label{eq:Z-UV2}
\end{align}
where the terms not given explicitly include 
subleading contributions from replacing $G(z)$ and $Z(z)$ by $G(y)$ and $Z(y)$,
and all contributions for $y<x$. Up to the latter,
the result for the tadpole is complete, because this loop integral,
the first integral on the right, does not depend on the external
momentum in the first place. 
Assuming constant $G(y)$ and $Z(y)$ one can see that 
the quadratically divergent terms are proportional to $\zeta-4$ as
they must.
The modifications to these terms introduced by using the
one-loop resummed forms for the dressing functions given in
Eqs.~(\ref{eq:G-UV}) and (\ref{eq:Z-UV}) will be considered below. 

Hence, choosing $\zeta=4$ is one possibility to get rid of spurious
divergences as observed by Brown and Pennington \cite{Brown:1988bn}. 
In particular, then the tadpole does not contribute at all.
In fact, this was one motivation to drop it in most previous studies.
However, since we work in the Landau gauge, we only need
transverse Green's functions, the subset of which  
closes among itself \cite{Fischer:2008uz}. The choice $\zeta=4$, on
the other hand, projects on the longitudinal parts of the
vertices also, and it introduces a spurious dependence on the 
additional gauge parameter $\eta$ which should not be there in
Landau gauge, requiring  $\zeta=1$ \cite{Lerche:2002ep}. Therefore,  
we use the transverse projector $P^{\zeta=1}_{\mu\nu}(p)$ from now on.

A number of proposals to remove the quadratically divergent parts
exist in the literature. We summarize them in a probably still incomplete
list as follows here: 
\begin{enumerate}

 \item \label{en:Brown_Pennington} Using the Brown-Pennington
   projector $P^{\zeta=4}_{\mu\nu}(p)$ gets rid of the divergent terms
   directly but introduces the ambiguity relating to $\eta$ 
   as discussed above.

 \item \label{en:integrands} Modifications of the integrand(s)
   \cite{Fischer:2002eq,Fischer:2002hn,Maas:2005hs,Cucchieri:2007ta,Huber:2012kd}: 
   The corresponding spurious terms can simply be subtracted from the
   integrands. Since the problem originates from the region of large
   loop momenta, the extra subtraction terms needed to achieve this 
   can be multiplied by a damping factor to avoid an influence on
   lower momentum regions at the expense of an additional
   parameter. Without such a damping factor, the 
   subtraction will in general also affect the IR. In this case it is
   advantageous to subtract the divergences for the loops in one
   integrand, if that particular loop is IR subleading
   \cite{Fischer:2002eq,Fischer:2002hn}.  
 
 \item \label{en:vertices} Modifications of the vertices
   \cite{Fischer:2008uz}: Instead of the integrands one can also modify
   the vertices as a technical trick in a way which does not
   reflect their behavior as expected from perturbation theory. To avoid an influence on the IR behavior, the 
   corresponding terms are again damped at low momenta, see also
   \cite{Maas:2011se}. 

 \item \label{en:fitting} Fitting the coefficient of the divergent
   part \cite{Fischer:2005en}: For dimensional reasons the
   quadratically divergent part is proportional to $1/p^2$. Fitting the
   coefficient of this term allows to subtract the corresponding mass-like term
   on the right-hand side which also gets rid of any potential tadpole
   contribution. This is implemented most easily for the scaling
   solution for which mass-like terms are IR subleading.   
 
\item \label{en:mass} Additional counter terms: In order to match decoupling
  solutions to lattice data a simple mass counter term has also been
  introduced \cite{Meyers:2014iwa} so that the $1/p^2$ terms are not
  subtracted but fixed by an additional condition hence again
  introducing an additional parameter. In two dimensions, on the
  other hand, the spurious divergences are logarithmic
  and can be subtracted via a kind of MOM scheme
  \cite{Huber:2012zj}.

 \item \label{en:dimreg} Dimensional regularization: There are no
   spurious divergences in analytic calculations using dimensional
   regularization by definition. Numerically it is difficult to realize
   \cite{Schreiber:1998ht,Gusynin:1998se}. It has so far only been
   implemented for logarithmic divergences and it is to our
   knowledge still not known yet how to numerically handle 
   power law divergences \cite{Phillips:1999bf}.
  
 \item \label{en:seagull} Seagull identities \cite{Aguilar:2009ke}:
   These identities were originally derived in dimensional
   regularization but they are also used within the PT-BFM framework
   \cite{Binosi:2009qm} with a momentum cutoff, e.g.,
   \cite{Binosi:2012sj,Aguilar:2013vaa}. Their use requires the
   vertices to have a special form so that the divergent parts of the
   individual integrals cancel via the seagull identities. 
  
\end{enumerate}

The general observation here is that most of the more practical
methods that have been implemented numerically are not unambiguous
because they involve a new a priori undetermined parameter.
This is the case, for example, when damping functions  are introduced
to subtract the integrands or via modifying the vertices according to
methods \ref{en:integrands} and \ref{en:vertices}. While one can
choose a value in the region of least sensitivity, some residual
dependence  on the damping parameter will always be left as the
price for cutoff independence in these schemes. This is avoided 
in method \ref{en:fitting} where one fits the coefficient in the 
mass-like $1/p^2$ term on the right-hand side of the DSE to subtract
it completely. However, this coefficient is in general not given by  
quadratically divergent contributions alone but can contain a finite
part which is then subtracted as well. Considering decoupling
solutions it is evident that such a finite part must exist due to the
massive behavior of the gluon propagator in the IR, i.e.,
$Z(p^2)\propto p^2/M^2 $ for low $p^2$. 
Also for the scaling solution, however, where it is IR as
well as UV subleading in the gluon DSE, we tested explicitly that
it can contribute beyond perturbation theory, as a dimension two
condensate contribution in the operator product expansion of the gluon
propagator, although we generally observe that its overall effects are
very small. Instead of fixing the finite part of the mass-like term 
by hand, as in Ref. \cite{Meyers:2014iwa} to match lattice data, we
will focus on disentangling perturbative from non-perturbative
contributions below in order to make sure that one subtracts only the
perturbative ones when removing spurious divergences.  

We emphasize, however, that as far as results are available that can be
compared, all methods yield qualitatively similar solutions, with
deviations that are still within the other systematic uncertainties  
at present. Given the current progress with enlarged truncations  
and quantitatively improving results especially for the gluon
propagator, on the other hand, a properly perturbative subtraction
method for spurious divergences should soon pay off as well.

\section{Subtraction of spurious divergences}
\label{sec:sub_procedure}

In the previous section we have exposed the spurious
divergences in the gluon propagator DSE by analyzing the UV behavior
of the integrands. This was not new, of course, but it was also used
to devise corresponding subtraction terms in the integrands or via the
vertices according to methods \ref{en:integrands} and
\ref{en:vertices}. The integrals over the subtraction terms used to
extend over all momenta and thus also affect the nonperturbative
region. This is not necessary, however, as will be seen explicitly
below. 

We again start from \eref{eq:Z-UV2}, set $\zeta=1$ and consider only
the spurious terms: 
\begin{align}
 \left(\frac{1}{Z_{UV}(p^2)}\right)_\mathrm{spur}\rightarrow &\, Z_4\frac{9N_c\,g^2}{64\pi^2}\frac{1}{x}\,\int_x^{\Lambda^2} dy Z_{UV}(y)\nnnl
 &+\frac{N_c\,g^2}{64\pi^2}\,\int_x^{\LL} dy \frac{1}{ x} G_{UV}(y)^2 -\frac{3\,N_c g^2}{32 \pi^2}\int_x^{\LL} dy \frac{1 }{ x}G_{UV}(y)^{2\alpha} Z_{UV}(y)^{2+2\beta}\nnnl
  =&\, Z_4\frac{9N_c\,g^2}{64\pi^2}\frac{Z(s)}{x}\,\int_x^{\Lambda^2} dy \left(\omega \,t_y\right)^{\gamma}\nnnl
 &+\frac{N_c g^2}{64 \pi^2}\frac{1}{ p^2}\left(G(s)^2-6G(s)^{2\alpha}Z(s)^{2+2\beta}\right)\int_x^{\LL} dy \left(\omega \,t_y\right)^{2\delta},\label{eq:Z_UV3}
\end{align}
where again $x=p^2$, $y=q^2$ and $t_y=\ln(y/\LQ)$. The logarithmic
divergences are handled separately, specifically by using a subtracted
equation. For now we suppress the corresponding extra terms. The
expression in \eref{eq:Z_UV3} depends on the external momentum only
via the trivial factor $1/p^2$. For the renormalized DSE, this
factor is modified to $1/p^2-1/p_0^2$, where $p_0^2=\mu^2 $ is the
subtraction point.   

\begin{figure}[tb]
 \includegraphics[width=0.49\textwidth]{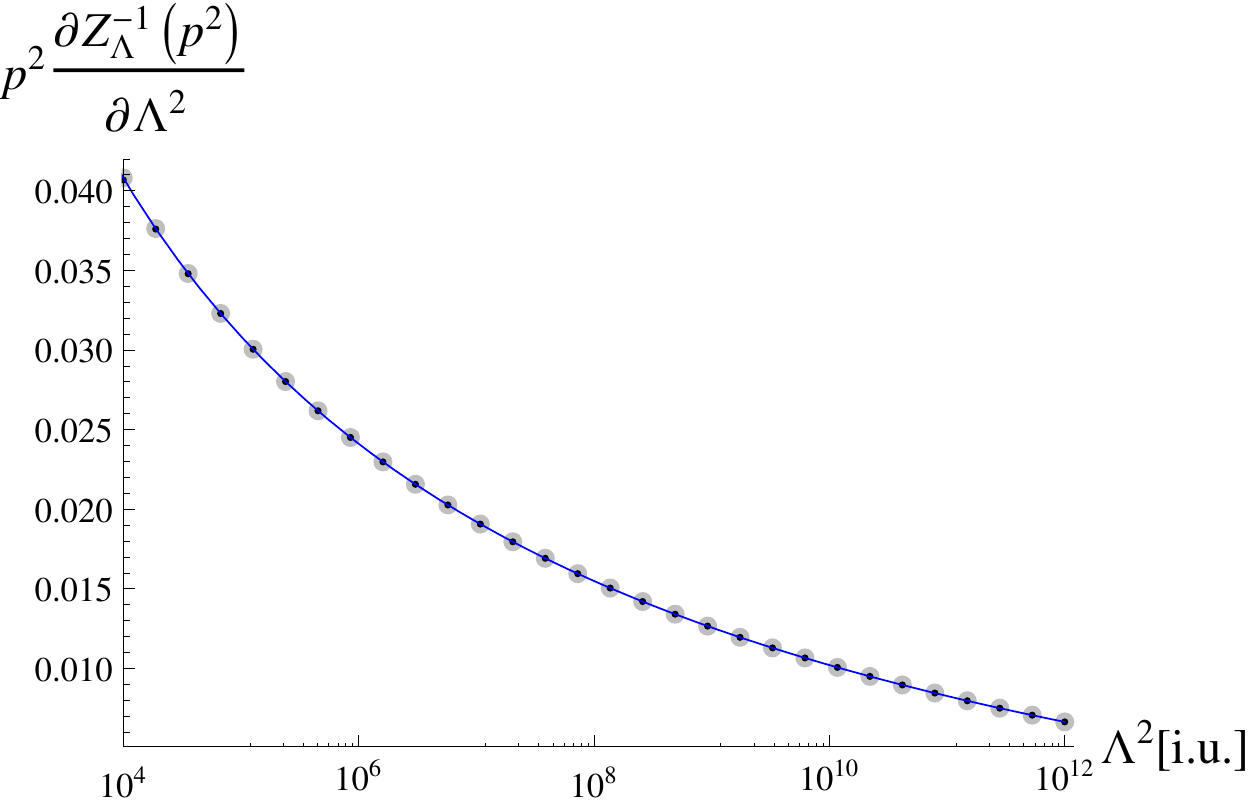}
 \includegraphics[width=0.49\textwidth]{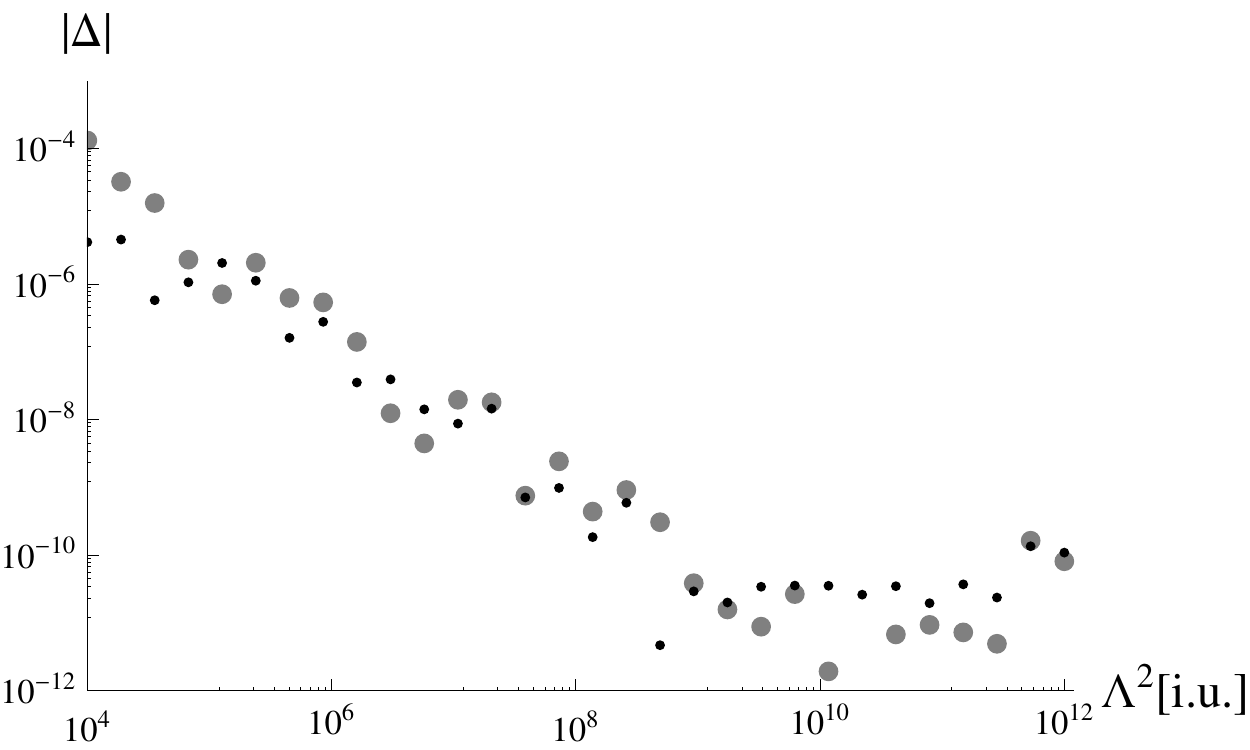}
 \caption{\label{fig:prop_dressings_comp_mod_cutoff}
 \textit{Left:} Cutoff dependence of the full solution
 $Z_\Lambda(p^2)$. The big gray (small black) dots represent results
 for the largest (smallest) external momentum used in the 
 calculations. The solid line shows the corresponding analytic expression
 from \eref{eq:Z_spur_diff}. The cutoff $\Lambda$ is given in arbitrary internal units.
 \textit{Right:} Absolute values of differences $\Delta$ in the
 left plot between smallest and largest momentum results (thick, gray dots),
 as well as numerical (smallest momentum) result and analytic
 expression (thin, black dots).} 
\end{figure}

In deriving this form of the quadratically divergent contributions 
we have replaced the mixed momentum $p+q$ in dressing functions by the
loop momentum $q$, for $q$ much larger than $p$, and it is assumed
that $s$ is sufficiently large as well so that we can use
Eqs.~(\ref{eq:G-UV}) and (\ref{eq:Z-UV}) for the leading UV behavior
of $G$ and $Z$. It turns out that this approximation is justified
extremely well, as can be verified by comparing the numerical
derivative with respect to the UV cutoff of the \textit{full}
solution $Z_\Lambda(p^2) $ of the renormalized DSE with the
corresponding result from the analytic expression above, which takes
the simple form: 
\begin{align}\label{eq:Z_spur_diff}
p^2 \frac{\partial Z_\mathrm{spur}^{-1}(p^2)}{\partial \Lambda^2} = b_{tad}\,
 (\omega\,t_{\Lambda})^{\gamma} + b \,(\omega\,t_{\Lambda})^{2\de},  
\end{align}
with $t_\Lambda=\ln\left(\LL/\LQ\right)$. The coefficients $b$ and $b_{tad}$ are given by
\begin{equation}
 b = \frac{N_c g^2}{64
   \pi^2}\left(G(s)^2-6G(s)^{2\alpha}Z(s)^{2+2\beta}\right)\, , \;\;
 \mbox{and} \quad 
 b_{tad} =Z_4\frac{9N_c\,g^2}{64\pi^2}\, Z(s)\,.  \label{eq:bbt}
\end{equation}
We also recall that
\begin{align}
 \omega &=\beta_{0}g^2(s)=\frac{11N_c \alpha(s)}{12\pi},\label{eq:w}
\end{align}
where the coupling $\alpha(s)$ at the large reference scale $s$ is
related to $\alpha(\mu)=g^2/4\pi$  at the renormalization point $\mu$,
in the {MiniMOM} scheme 
\cite{vonSmekal:2009ae}, via
 \begin{align}
 \alpha(s)=\alpha(\mu^2)Z(s) G(s)^2 ,
\end{align}
but other schemes may be used as well.  
Calculating the derivative has two advantage. First, it avoids large
numbers and it becomes easier to expose the general form of the
expression. Secondly, all cutoff independent terms drop out. Since the
logarithmic cutoff dependence has already been taken care of 
by momentum subtraction, the remaining expression only contains
contributions from spurious divergences. A comparison of the
analytic result in \eref{eq:Z_spur_diff} with two numerical results
from $Z_\Lambda(p^2)$ is shown in
\fref{fig:prop_dressings_comp_mod_cutoff}.  
The latter two are those for the minimal and the maximal external
momenta used in the calculation. The good agreement is demonstrated
in the right plot of \fref{fig:prop_dressings_comp_mod_cutoff}, where
the difference between the two is shown. Hence, 
spurious quadratic divergences indeed depend on the external momentum only
via the prefactor $1/p^2$, i.e., there is no evidence of a nontrivial
momentum dependence in these terms. The solid line in the left plot of 
\fref{fig:prop_dressings_comp_mod_cutoff} represents the analytic
result from \eref{eq:Z_spur_diff} which coincides with the numerical
results, again within numerical accuracy as also demonstrated on the
right. Consequently we conclude that we have correctly identified the
source of the spurious divergences in \eref{eq:Z_UV3} to be of purely
perturbative origin. 

To remove the spurious terms it is sufficient to subtract the
indefinite integral of \eref{eq:Z_spur_diff} over $\Lambda^2$.
The integration constant then again introduces an arbitrary mass term,
however. In other words, if we replace the lower integration bound
$x=p^2$ in \eref{eq:Z_UV3} by an arbitrary constant $x_1$ in the
necessary subtraction term, we have traded the momentum dependence there for
a mass term.  Our goal, however, is to
implement this subtraction in a minimal way, in particular, we want to avoid
adding an ad hoc mass term. This will be achieved by choosing the special
$x_1 $ at which all cutoff independent terms vanish in the subtraction term identically.

Appendix~\ref{sec:quad_div_term} contains the details of integrating 
 \eref{eq:Z_spur_diff} from a general lower bound $x_1$
 to the UV cutoff. The two necessary integrals only differ in their
exponent so we only give the solution for $\gamma$ here.  With $t_1=\ln\left(x_1/\LQ\right)$ it reads
\begin{align}
I(\LL,\gamma)-I(x_1,\gamma) &=\int_{x_1}^{\LL} dy \left(\omega\,
t_{y}\right)^{\gamma}= \LQ (-\omega)^{\gamma}\, \big( \Gamma\left(1+\gamma, -t_{\Lambda}\right) - \Gamma\left(1+\gamma,-t_{1}\right)\big)
\end{align}
The definition of the incomplete Gamma function $\Gamma(a,z)$ is given
in \eref{eq:inc_gamma}. 
The lower bound $x_1$ is now fixed such that the constant contribution
vanishes, which leads to $t_{1}=0$ and hence $x_1=
\Lambda_\mathrm{QCD}^2$. In other words, we integrate down to the
Landau pole of the perturbative running coupling. The subtraction
coefficient of the spurious $1/p^2$ terms then reads   
\begin{align}\label{eq:C_sub}
C_\mathrm{sub}&:=b \left(I(\LL,2\delta)- I(\LQ,2\delta)\right)+b_{tad}\left(I(\LL,\gamma)-I(\Lambda_\mathrm{QCD}^2,\gamma)\right)\nnnl
&=\LQ\left(b\,\omega^{2\delta}\sum_{n=0}^{\infty}\frac{(t_{\Lambda})^{1+2\delta+n}}{n!(1+2\delta+n)}+b_{tad}\,\omega^{\gamma}\sum_{n=0}^{\infty}\frac{(t_{\Lambda})^{1+\gamma+n}}{n!(1+\gamma+n)}\right)
\end{align}
where we have used the series expansion of the incomplete Gamma
function in \eref{eq:inc_gamma_ser}. In this form it is numerically
easy to calculate. An added bonus of choosing the perturbative Landau
pole as the lower integration bound in the subtraction term is that
the details of the lower momentum behavior of the corresponding
integrands do not matter at all as long as the contributions to the
integrals from the lower bound vanish. As an example we discuss an
alternative parametrization of the perturbative 
propagators, with the same leading perturbative behavior in the
UV but with the Landau pole and $x_1$ both shifted to zero in
Appendix~\ref{sec:alt_UV_parametrization}. 
The result is the same.

Another reason that makes this choice a natural one is the following: Spurious divergent terms appear already for a purely perturbative treatment and have to be handled there as well. Within such a calculation, a constant contribution, as arising from a contribution from the lower boundary, would correspond to a mass term for the gluon, which is of course not allowed. Since the cutoff dependence is the same for all external momenta, as demonstrated above, we can use the perturbative prescription in the nonperturbative regime as well. It should be noted that the massive behavior (in the sense of a screening mass) at low momenta for the decoupling solution of the gluon propagator is of purely nonperturbative origin.

The suggested procedure to subtract spurious divergences is summarized as follows:
\begin{enumerate}

 \item Calculate $b$, $b_{tad}$ and $\omega$ from eqs.~(\ref{eq:bbt})~and~(\ref{eq:w}).

 \item Calculate the subtraction coefficient $C_\mathrm{sub}$ from \eref{eq:C_sub}.

 \item Subtract the spurious divergences via
  \begin{align}\label{eq:sub_prescription}
  Z(p^2)^{-1}:=Z_\Lambda(p^2)^{-1}-C_\mathrm{sub}\left(\frac{1}{p^2}-\frac{1}{p_0^2}\right)
  \end{align}
  where $Z_\Lambda(p^2)^{-1}$ corresponds to the calculated right-hand side of the gluon propagator DSE.

\end{enumerate}

\begin{figure}[tb]
 \includegraphics[width=0.49\textwidth]{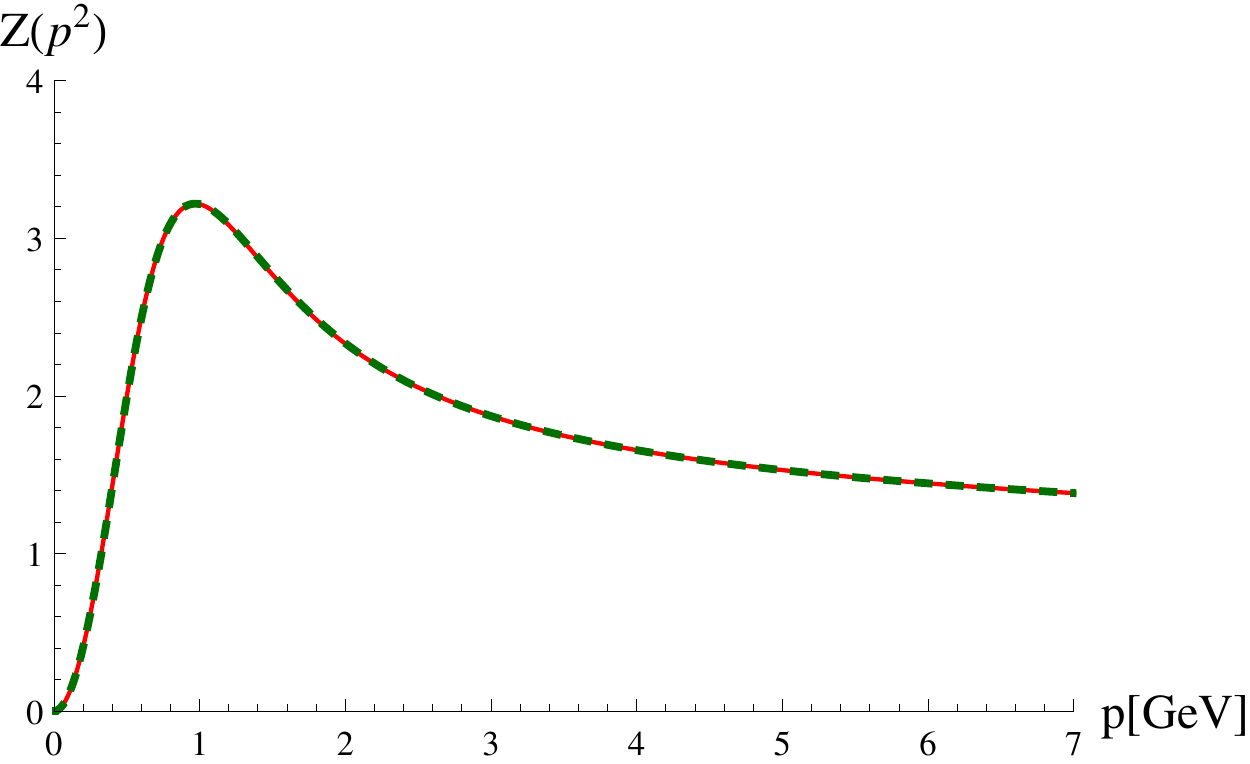}
 \caption{\label{fig:prop_dressings_diff_cutoffs}The gluon dressing function for two cutoffs: $7747$~GeV$^2$ (red, continuous line) and twice that magnitude (green, dashed line).
}
\end{figure}

In \fref{fig:prop_dressings_diff_cutoffs} we show results obtained
with this procedure. The two lines, calculated with two different
UV cutoffs, agree with each other at the level of the numerical
precision. Another check was to vary the routing of the external
momentum through the loop diagrams. For this we transformed the loop
momentum as $q\rightarrow q+\lambda\, p$ in \eref{eq:Z} and tested
several values of $0\leq\lambda\leq1$. In all calculations the results
were the same within the precision seen in
\fref{fig:prop_dressings_diff_cutoffs}. 

\section{Effects on the gluon propagator dressing}
\label{sec:effects_gluon}

We will now apply the method discussed above to the coupled system of propagators of Landau gauge Yang-Mills theory. For the three-gluon vertex dressing the following model is employed~ \cite{Huber:2012kd}:
\begin{align}\label{eq:3g-model}
D^{A^3}(p,q,k)&=G\left(\frac{x+y+z}{2}\right)^{\alpha}Z\left(\frac{x+y+z}{2}\right)^{\beta}-G(x+y+z)^{3}(f^{3g}(x)f^{3g}(y)f^{3g}(z))^4,
\end{align}
where $f^{3g}(x)$ is a damping function given by
\begin{align}
f^{3g}(x)&:=\frac{\Lambda^2_{3g}}{\Lambda_{3g}^2+x}.
\end{align}
The first part of the three-gluon vertex dressing was already used in the UV analysis in Sec.~\ref{sec:quadDivs} and describes the behavior according to one-loop resummed perturbation theory. The second part was introduced to account for the nonperturbative behavior of the vertex dressing which becomes negative in the IR \cite{Huber:2012kd,Pelaez:2013cpa,Aguilar:2013vaa,Blum:2014gna,Eichmann:2014xya}. The logarithmic IR divergence found in several approaches \cite{Pelaez:2013cpa,Aguilar:2013vaa,Blum:2014gna,Eichmann:2014xya} is not implemented. Since the vertex is always multiplied by gluon dressing functions which are suppressed in the IR this is not of relevance here. The parameter $h_{IR}$ is set to $-1$ and $\Lambda_{3g}$ can be varied. For the calculations the programs \textit{DoFun} \cite{Huber:2011qr,Alkofer:2008nt} and \textit{CrasyDSE} \cite{Huber:2011xc} were employed.

\begin{figure}[tb]
 \includegraphics[width=0.49\textwidth]{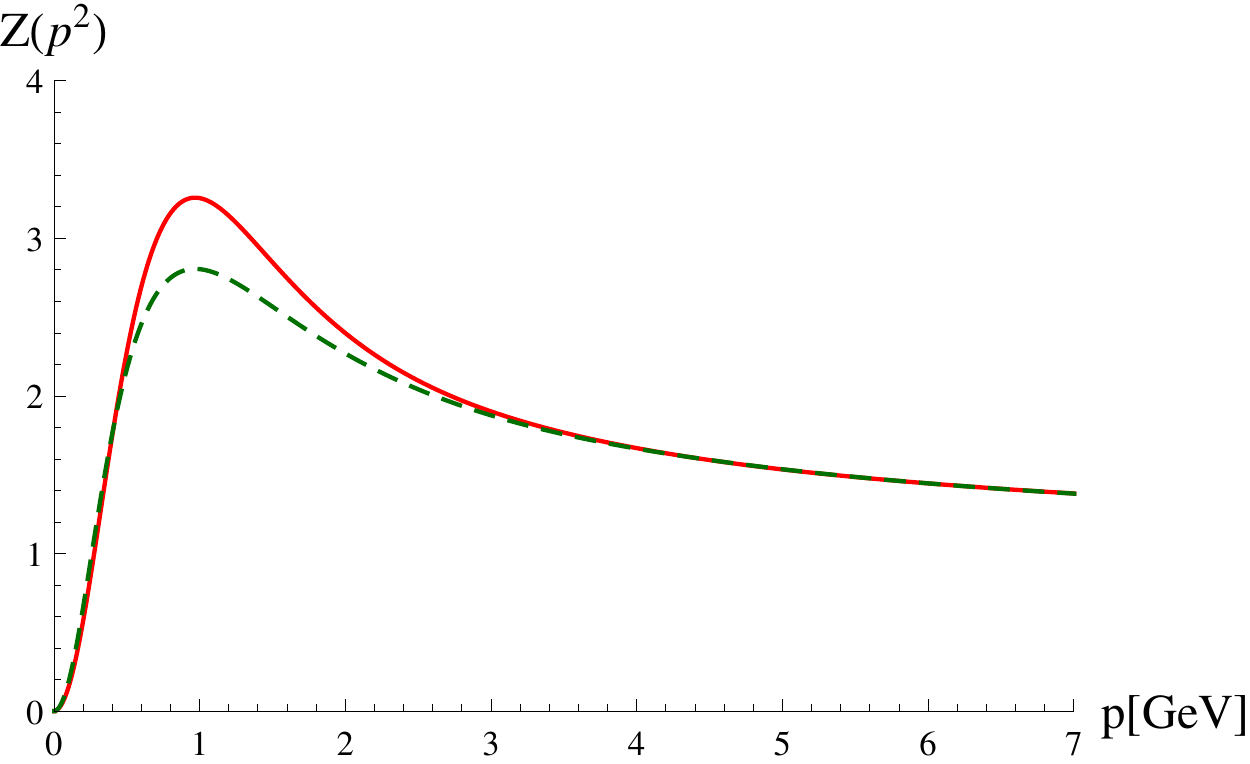}
 \includegraphics[width=0.49\textwidth]{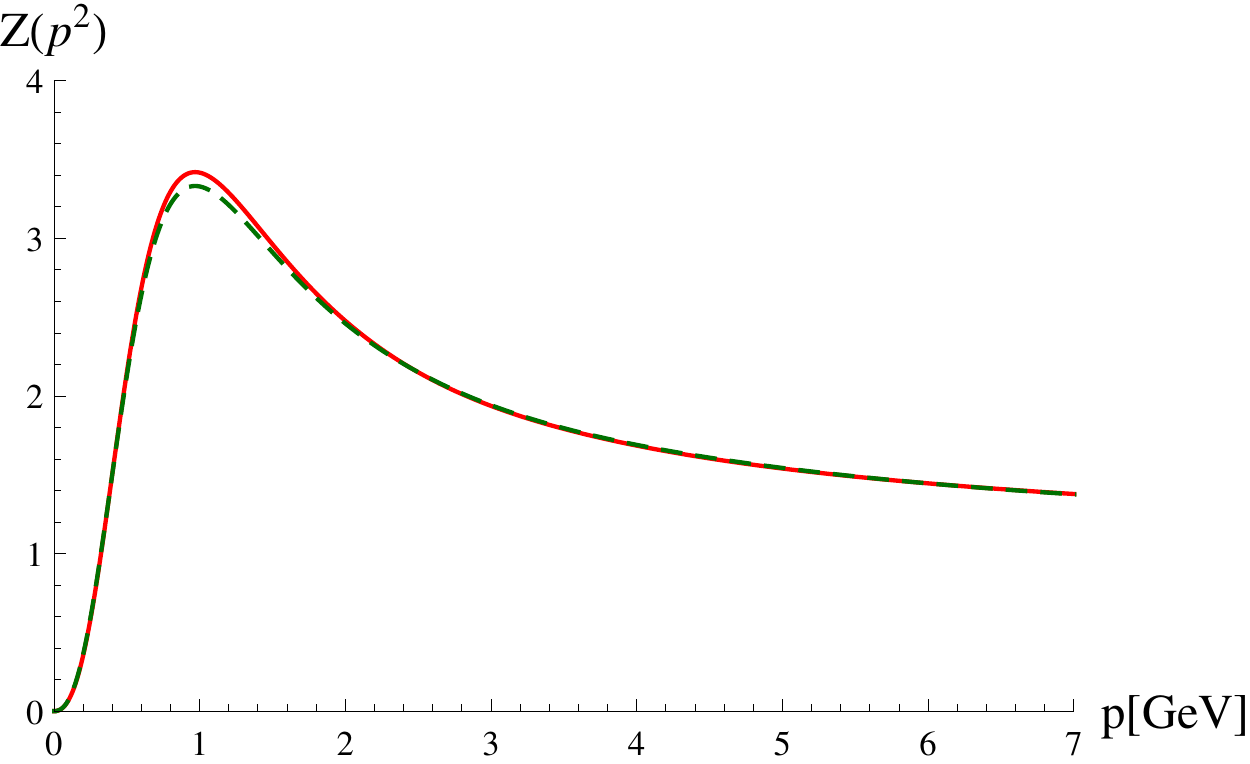}
 \caption{\label{fig:prop_dressings_diff_methods}The gluon dressing functions for decoupling/scaling (\textit{left/right}) for different subtraction methods and with $\Lambda_{3g}= 0.82$ GeV in the three-gluon vertex model of \eref{eq:3g-model}. For $D^{A^3}_{RG}$ \eref{eq:DAAA-UV} with full propagators was used. Green, dashed line: Subtraction in the gluon loop as in refs.~\cite{Fischer:2002eq,Fischer:2002hn}. Red, continuous line: Subtraction via \eref{eq:sub_prescription}.}
\end{figure}

\begin{figure}[tb]
 \includegraphics[width=0.49\textwidth]{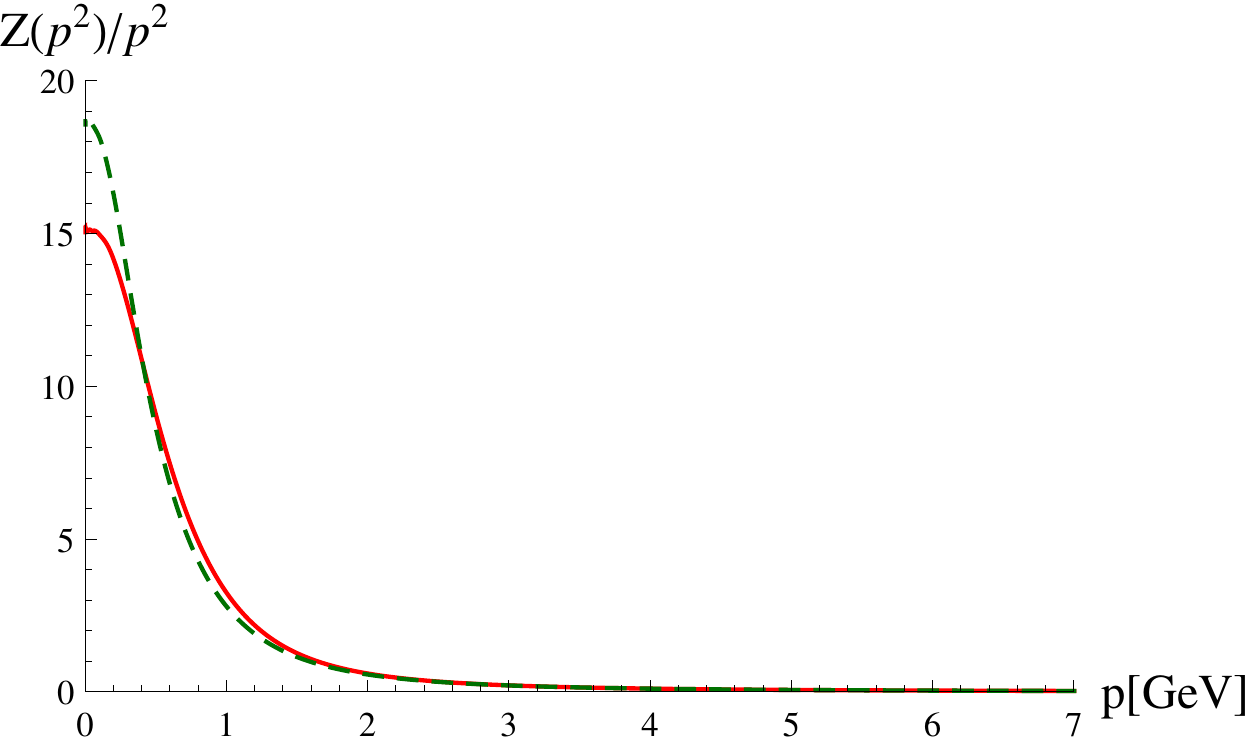}
 \includegraphics[width=0.49\textwidth]{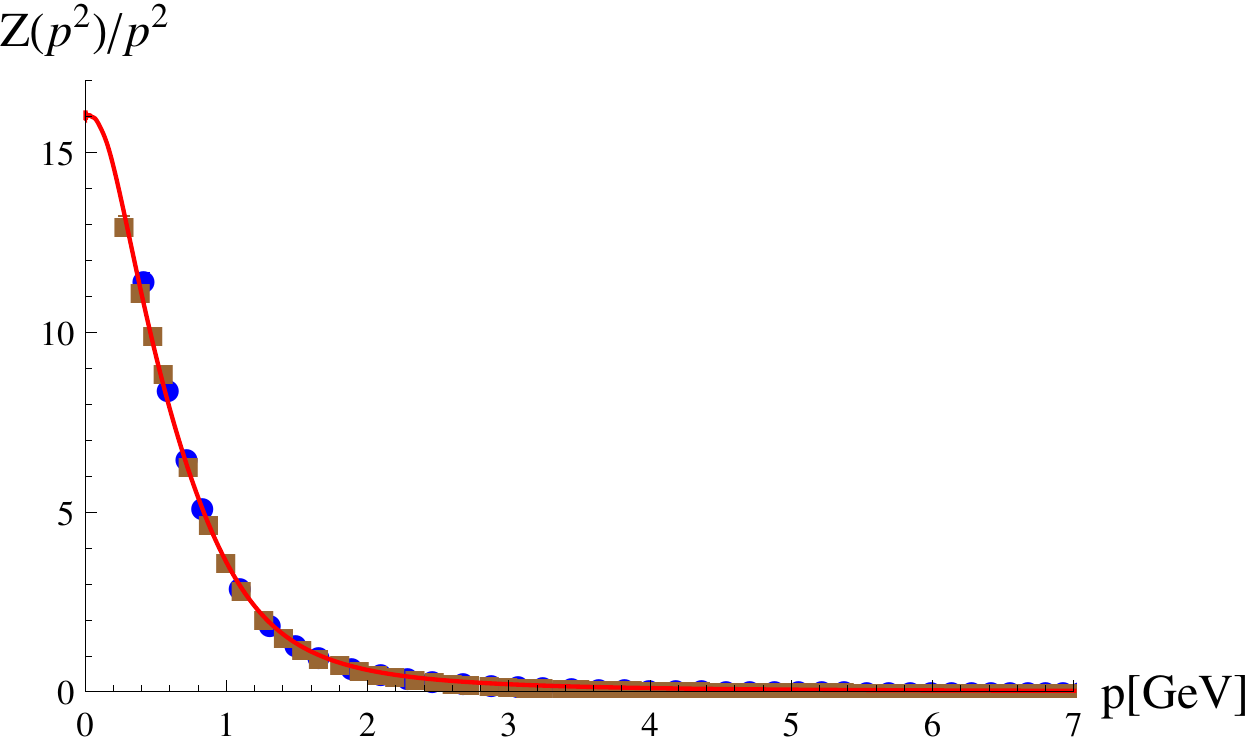}
 \caption{\label{fig:gl_props}\textit{Left/Right:} The gluon propagators corresponding to the results in Figs.~\ref{fig:prop_dressings_diff_methods} (left) and \ref{fig:prop_dressings_compL}.}
\end{figure}

\subsection{Comparison of subtraction methods}

First we illustrate to what extent results differ when using different methods of subtraction.
For easier separation of the individual effects we do not include the
tadpole diagram here yet. Its role is discussed separately below.
In \fref{fig:prop_dressings_diff_methods} the dashed, green line was
obtained by using a variant of method \ref{en:integrands}: The
divergent terms are subtracted in the gluon loop without a damping
function \cite{Fischer:2002eq,Fischer:2002hn}. The continuous, red
line was obtained from the method proposed here. The maximal
difference occurs at momenta around $1$~GeV. Its magnitude is
different for scaling and decoupling solutions, the reason being most
likely that the (finite) $1/p^2$ terms are IR subleading for
scaling. Hence the difference is smaller for scaling than for
decoupling where there is a clear deviation that is manifest for the
dressing function in the mid-momentum regime. Plotting the propagator
instead of the dressing function confirms that also the IR is affected, see
\fref{fig:gl_props}, which means that the gluon screening mass $M^2
:=D(0)^{-1}$ obtained from both methods is different. This is a
natural consequence of the fact that the gluon screening mass arises
from terms that behave like $1/p^2$ on the right-hand side of the
gluon DSE, which is the same momentum dependence as that of the
spurious divergences. 
Although the difference between the two methods is quite large, the
spread is still within the error introduced by the modeled vertices,
see, e.g., Refs.~\cite{Pennington:2011xs,Huber:2012kd}.

\subsection{Renormalization group improvement}

\begin{figure}[tb]
 \includegraphics[width=0.49\textwidth]{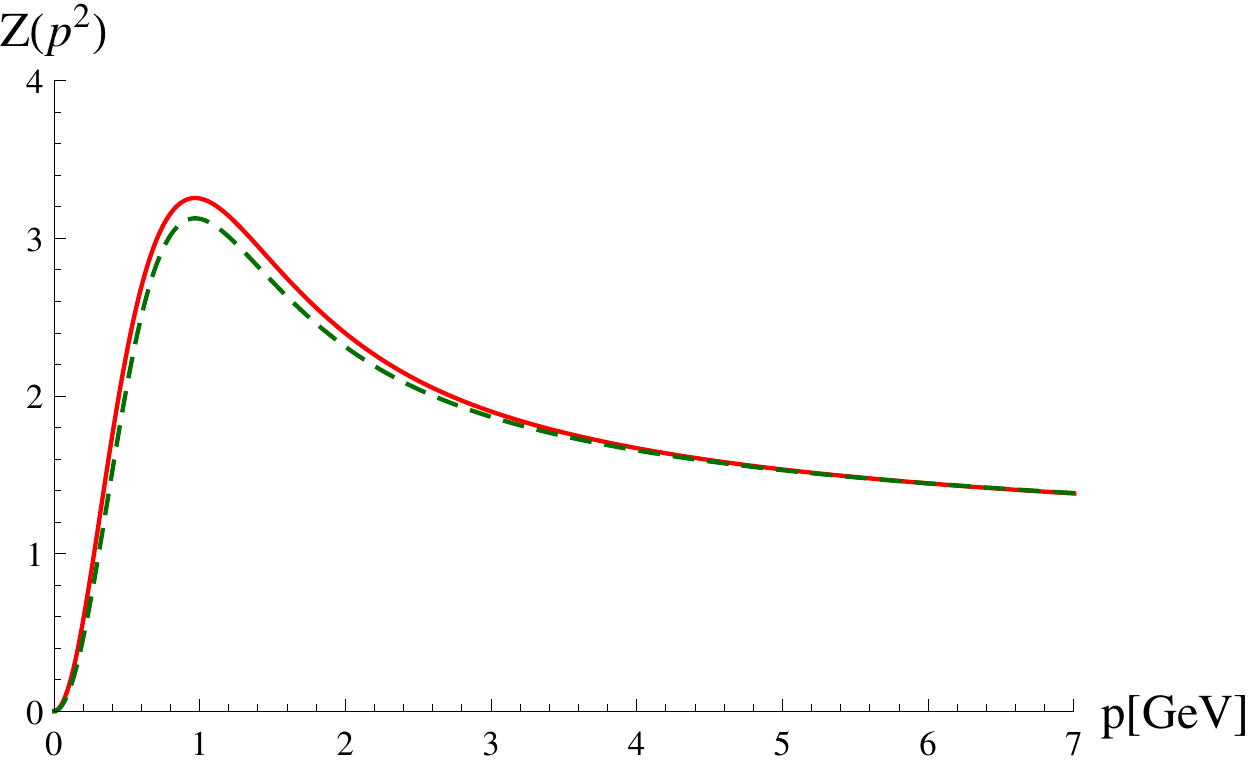}
 \includegraphics[width=0.49\textwidth]{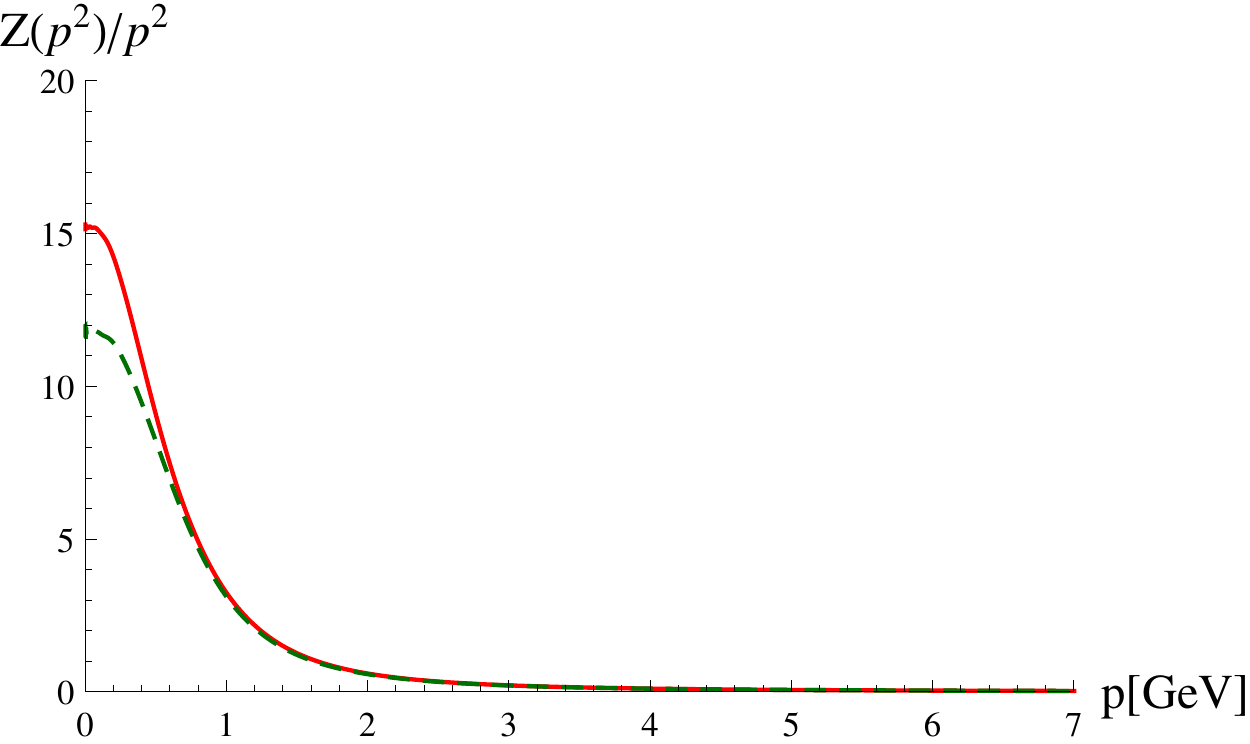}
 \caption{\label{fig:RGI}The gluon dressing function and propagator calculated with $D^{A^3}_{RG}$ from \eref{eq:DAAA-UV} with full propagators (red, continuous) and with $D^{A^3}_{RG}$ from \eref{eq:DAAA-RG} (green, dashed).}
\end{figure}

It is known that the choice of the RG improvement has some influence on the results for the dressing function. Specifically for the three-gluon vertex it was shown that the IR is affected rather strongly as the position of its zero crossing is sensitive to this choice \cite{Eichmann:2014xya}. Here we propose an alternative expression for $D^{A^3}_{RG}(p^2,q^2,k^2)$ that eliminates certain ambiguities related to the behavior in the IR. For this we define the following dressing functions that obey the correct UV behavior and become unity in the IR:
\begin{align}
 G_{RG}(x):=G(s)\left(\omega\,\ln \left(a_{gh}+\frac{x}{\Lambda^2_\mathrm{QCD}} \right)\right)^\delta,\\
 Z_{RG}(x):=Z(s)\left(\omega\,\ln \left(a_{gl}+\frac{x}{\Lambda^2_\mathrm{QCD}} \right)\right)^\gamma,
\end{align}
with
\begin{align}
 a_{gl}=e^{\frac{Z(s)^{-1/\gamma}}{\omega}}, \quad
 a_{gh}=e^{\frac{G(s)^{-1/\delta}}{\omega}}.
\end{align}
$G_{RG}(x)$ and $Z_{RG}(x)$ are used to define a renormalization group improvement term that has the advantage over previous choices that it becomes unity in the IR and thus does not affect the results there. In particular, this expression is the same for decoupling and scaling solutions, because it only depends on the UV of the propagators. It is given by
\begin{align}
 D^{A^3}_{RG}(p^2,q^2,k^2)&= \frac{G_{RG}(\overline{p}^2)}{Z_{RG}(\overline{p}^2)}.\label{eq:DAAA-RG}
\end{align}
The specific form of this expression is motivated by the STI, $Z_1=Z_3/\widetilde{Z}_3$.
The influence of the choice of $D^{A^3}_{RG}(p^2,q^2,k^2)$ on the gluon propagator is shown in \fref{fig:RGI}, where we compare the solution obtained with the renormalization group improvement from \eref{eq:DAAA-RG} to the a solution where $D^{A^3}_{RG}(p^2,q^2,k^2)=D^{A^3}_{UV}(p^2,q^2,k^2)$ was used. While the difference in the midmomentum region is small, the IR is affected stronger. Note that $Z_4$ in the tadpole diagram is not replaced as it does not contribute to the anomalous dimension of the gluon propagator. Its value is determined via the STI $Z_4=Z_3/\widetilde{Z}_3^2$. Typical values in our calculations are $Z_3\approx3.98$ and $\widetilde{Z}_3\approx1.56$. 

\subsection{Tadpole contributions}

\begin{figure}[tb]
 \includegraphics[width=0.49\textwidth]{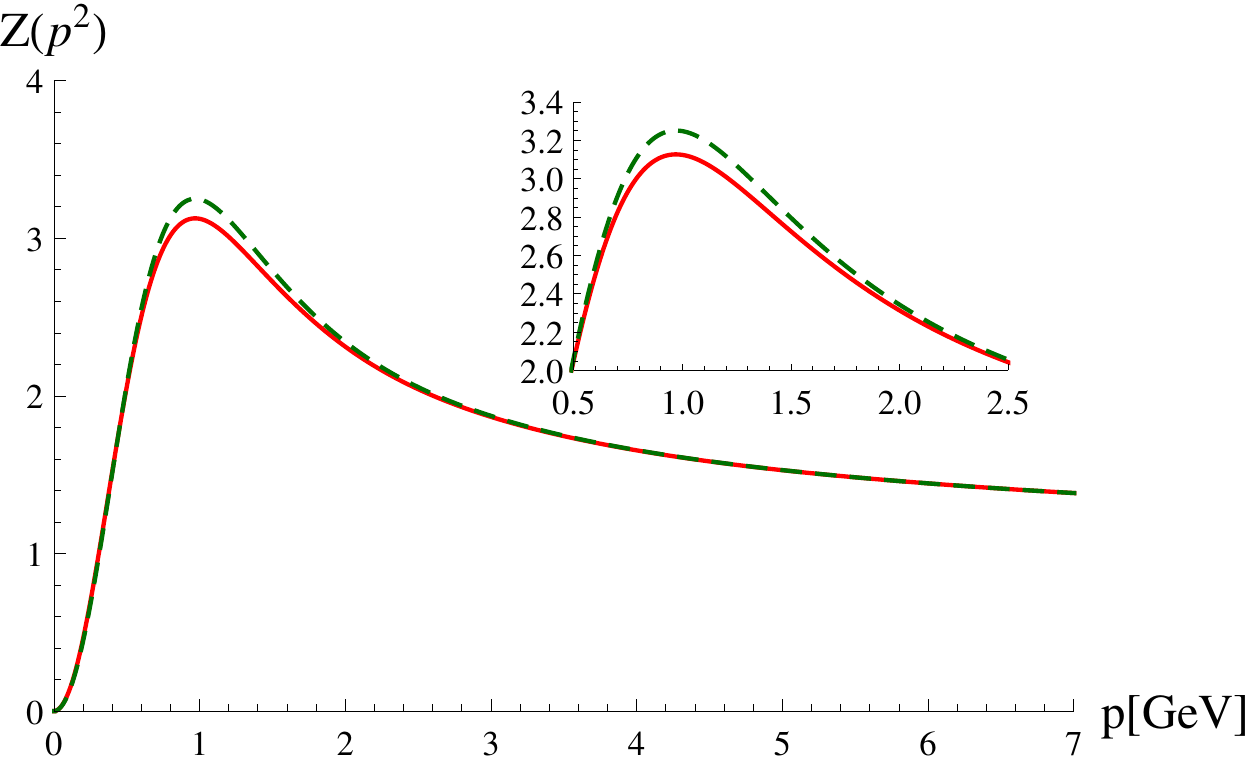}
 \includegraphics[width=0.49\textwidth]{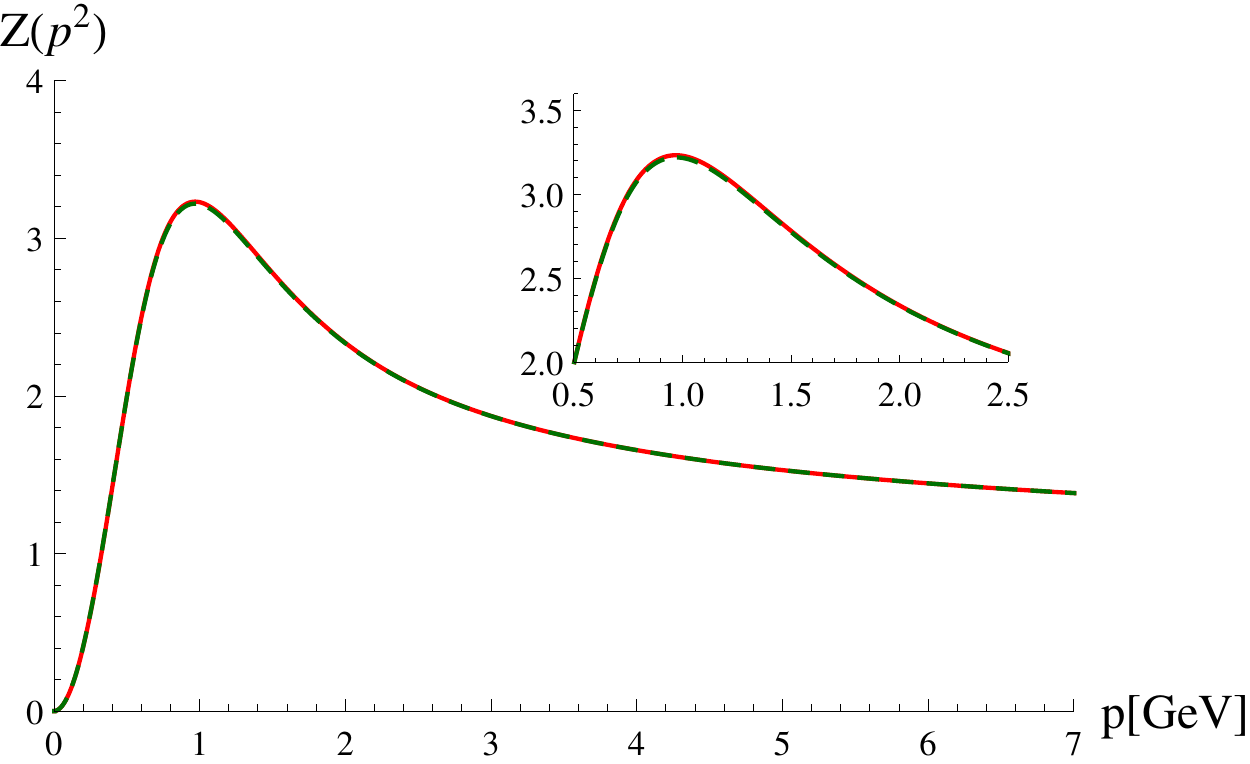}
 \caption{\label{fig:gl_tadpole}The gluon dressing function calculated without (red, continuous) and with (green, dashed) tadpole. For $D^{A^3}$ \eref{eq:3g-model} with $\Lambda_{3g}=0/1.7$~GeV (\textit{left}/\textit{right}) was used, for $D^{A^3}_{RG}$ \eref{eq:DAAA-RG}.}
\end{figure}

This diagram is typically dropped in numeric calculations as it is believed that it does not contribute to the gluon selfenergy. This relies on several arguments. First of all, in the UV analysis it turned out that the complete diagram is proportional to $\zeta-4$ and thus looks like a pure spurious divergence. Furthermore, in dimensional regularization this diagram is zero in perturbation theory. However, this does not exclude nonperturbative contributions. From the form of the tadpole integrand it becomes evident that subtracting the spurious divergences in our scheme leaves us with the integral over the difference between the perturbative and nonperturbative gluon propagators. Thus, as required by the absence of a gluon mass in perturbation theory, there is no perturbative contribution but only one from the nonperturbative regime.
We found that the magnitude of the subtracted tadpole integral depends on the details of the employed three-gluon vertex model. In \fref{fig:gl_tadpole} two examples are shown. We found cases, where the difference is negligible, but we also obtained solutions where differences of several percent were found. They only occur in the nonperturbative regime and cause, for example, a different height of the bump of the gluon dressing function and a different gluon screening mass.

\begin{figure}[tb]
 \includegraphics[width=0.49\textwidth]{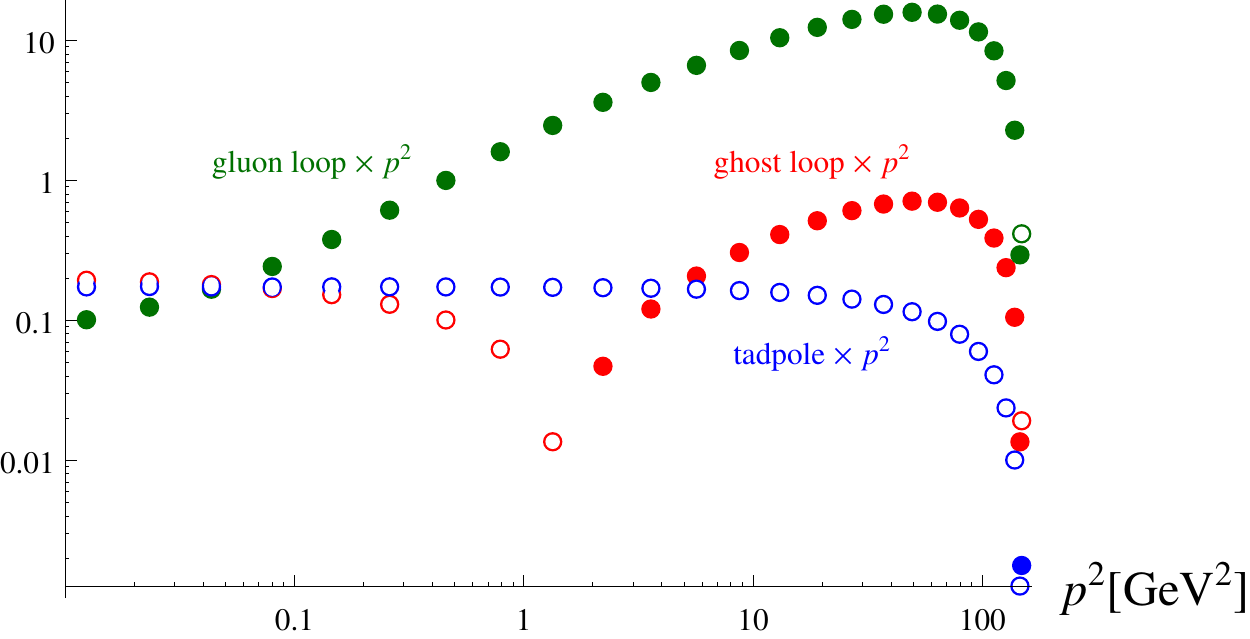}
 \includegraphics[width=0.49\textwidth]{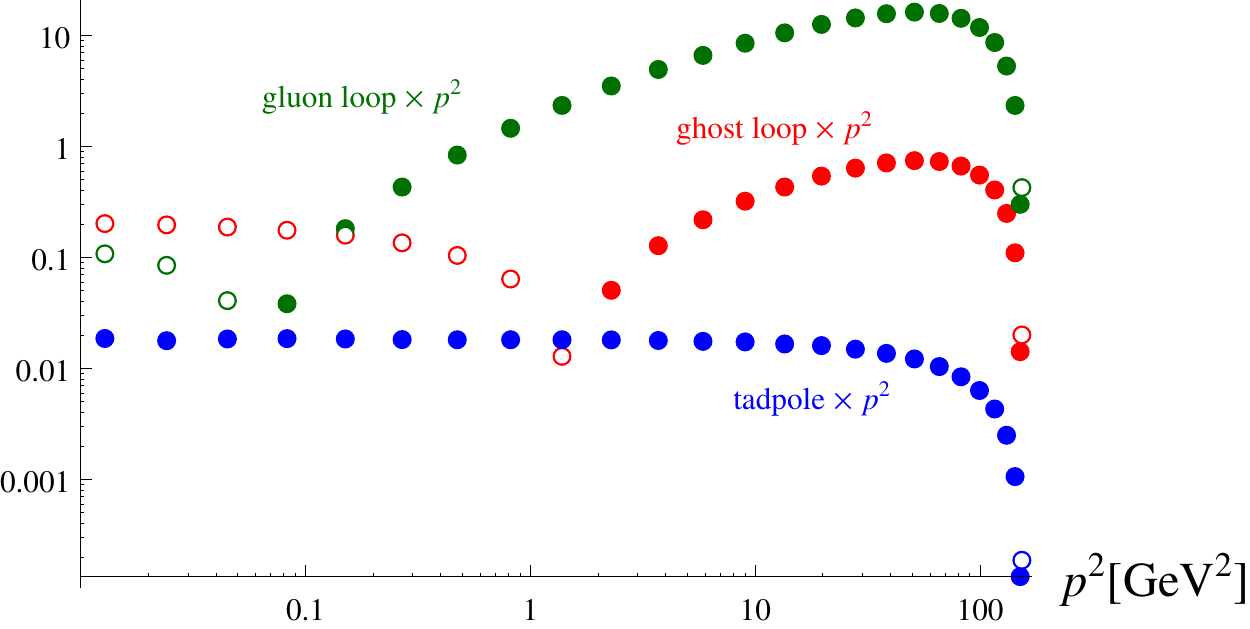}
 \caption{\label{fig:diagram_contributions}Individual contributions of the three one-loop diagrams in the gluon propagator DSE corresponding to the results in \fref{fig:gl_tadpole} multiplied by $p^2$. Full circles denote negative values. The second zero crossing is determined by the subtraction point.}
\end{figure}

The method proposed here has also the advantage that contributions from single diagrams can be disentangled. Thus the importance of the ghost and the gluon sectors can be considered separately. However, one should keep in mind that due to the nonlinearity of the gluon propagator DSE each part is important on its own for the final result albeit the magnitude of the contribution can be small. The individual contributions corresponding to the dressings in \fref{fig:gl_tadpole} are plotted in \fref{fig:diagram_contributions}. To better expose their momentum dependence they were multiplied by $p^2$. As expected the gluon loop dominates in the mid- and high-momentum regimes. The ghost loop becomes important for small momenta.

\subsection{Calculations with an optimized effective three-gluon vertex}

\begin{figure}[tb]
 \includegraphics[width=0.49\textwidth]{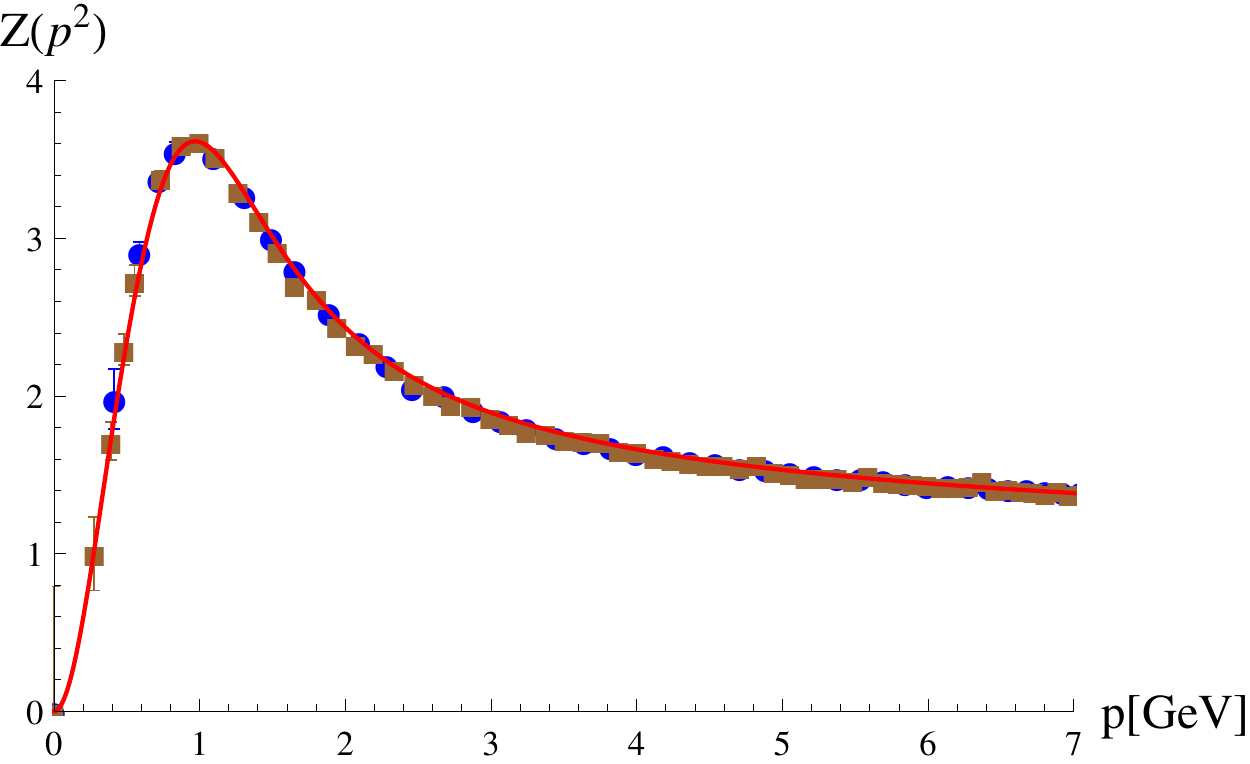}
 \includegraphics[width=0.49\textwidth]{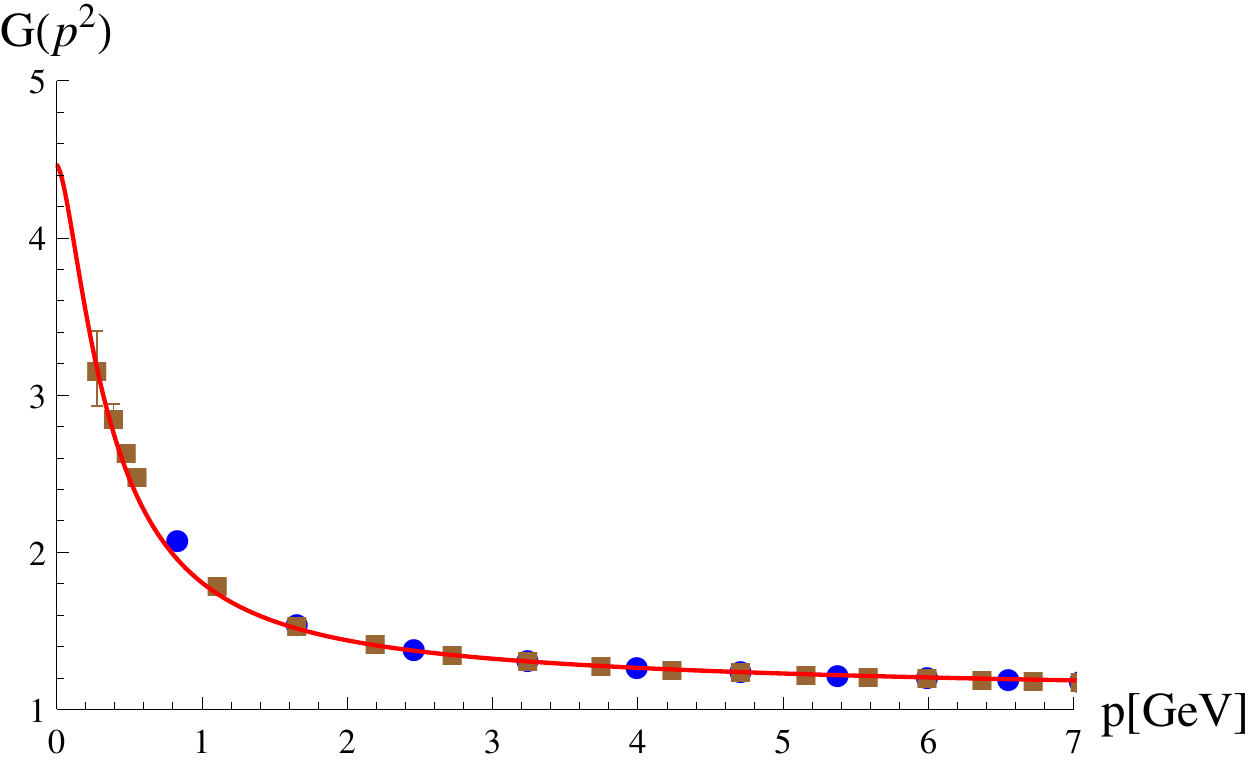}
 \caption{\label{fig:prop_dressings_compL}The gluon and ghost dressing functions with a bare ghost-gluon vertex and $\Lambda_{3g}=2.9$ GeV for the three-gluon vertex model of \eref{eq:3g-model} compared to lattice data \cite{Sternbeck:2006rd} (blue disks: $N=32$; brown squares: $N=48$; both $\beta=6$).}
\end{figure}

In ref.~\cite{Huber:2012kd} it was observed that the model for the three-gluon vertex can be used to effectively include contributions of the two-loop terms. The model with the best choice of parameters was called optimized effective three-gluon vertex. Since the midmomentum is affected by the subtraction method for the spurious divergences we repeat this calculation here, i.e., we drop the tadpole diagram and use \eref{eq:DAAA-UV} with full propagators for $D^{A^3}_{RG}$. This is justified, since the employed model for the three-gluon vertex including the RG improvement term effectively mimics the missing contributions. The results, which reproduce results from Monte-Carlo simulations very well, can then be used as input in other calculations. For the parameter $\Lambda_{3g}$ we found $2.9$~GeV, whereas in \cite{Huber:2012kd} it was $1.8$~GeV. The results for the dressing functions are compared to lattice data in \fref{fig:prop_dressings_compL}. The gluon propagator is depicted in the right plot of \fref{fig:gl_props}. These plots show that good agreement with lattice results is obtained. However, we stress that in a full calculation with the correct three-gluon vertex this can only be achieved when the tadpole and the two-loop diagrams are included. The contributions of the former were investigated here, while for the latter calculations showed that the squint diagram can yield sizable contributions  \cite{Bloch:2003yu,Meyers:2014iwa} and the sunset is strongly subleading \cite{Bloch:2003yu,Mader:2013ru,Meyers:2014iwa}.

\section{Summary and conclusions}
\label{sec:summary_conclusions}

Spurious divergences in the gluon propagator DSE are an obstacle that
needs proper handling. The various methods in the literature differ
from each other, but only to an extent that is within the error
expected from truncating the gluon propagator DSE. Thus, for
quantitative improvements a good understanding of spurious divergences
is necessary. As required, the regularization procedure we described
here removes any dependences on the cutoff beyond logarithmic
divergences. One parameter $x_1$ enters via the lower bound of an
integral. Since spurious divergences have their origin in the
perturbative regime, we argued that $x_1$ should be fixed such that
perturbatively no mass terms are introduced. As checks of our results we varied the cutoff and the momentum routing without any effect on the obtained dressing function of the gluon propagator. Also an alternative parametrization of the perturbative behavior that leads to a different value for $x_1$ was discussed and confirmed our results.

Our regularization prescription for spurious divergences is summarized at the end of Sec.~\ref{sec:quadDivs}. We illustrated its use by calculating the ghost and gluon propagators using an optimized effective three-gluon vertex which allowed us to reproduce lattice results rather accurately, see \fref{fig:prop_dressings_compL}. Furthermore, this method allows to obtain the nonperturbative contribution from the tadpole diagram. Its influence on the gluon dressing function can be up to a few percent, but its magnitude depends on details of the employed model for the three-gluon vertex. 

Note that while we described this procedure only for the Yang-Mills
sector of QCD explicitly, the inclusion of the quark propagator does
not lead to any new problems. With a suitable ansatz for the
quark-gluon vertex that respects the correct UV behavior, the formal
structure of the quark loop is the same as that of the ghost and gluon
loops as can be inferred from its UV analysis, see, e.g.,
\cite{Fischer:2003rp}. Thus the spurious term arising from the quark
loop is of the same form and can be subtracted as described here.

Dynamically including vertices, on the other hand, requires an
extension of this method. The reason is that the models employed here
respect the one-loop resummed behavior in the UV exactly and no higher
perturbative corrections are included. However, using calculated
vertices means that higher loop effects enter. It remains to be seen
if this can be taken into account analytically or if a numeric
approach, for example fitting to \eref{eq:Z_spur_diff}, is more
promising.

\section*{Acknowledgments}

We thank Christian S. Fischer and Markus Hopfer for useful discussions. We are particularly grateful to Richard Williams for discussions about subtraction method \ref{en:fitting} and a careful reading of the manuscript. 
This work was supported by the Helmholtz International Center for FAIR within the LOEWE program of the State of Hesse and the European Commission, FP7-PEOPLE-2009-RG No. 249203.

\appendix

\section{Calculation of the subtraction coefficient $C_\mathrm{sub}$}
\label{sec:quad_div_term}

To calculate $C_\mathrm{sub}$ the following integral has to be solved:
\begin{align}
\label{eq:I2}
 I(\Lambda^2,\gamma)-I(x_1,\gamma)=\int_{x_1}^{\LL} dy \left(\omega \, t_y\right )^{\gamma},
\end{align}
where $x_1$ is a cutoff independent value for the lower bound. Using $u(y)=- t_y$ as integration variable we obtain
\begin{align}
 I(\Lambda^2,\gamma)-I(x_1,\gamma) = -\LQ(-\omega)^{\gamma}\int_{u(x_1)}^{u(\LL)} du\,e^{-u}u^{\gamma}.
\end{align}
The result can be expressed in terms of the incomplete Gamma function, given by
\begin{align}\label{eq:inc_gamma}
\Gamma(a,z)=\int_z^\infty du\, u^{a-1}e^{-u}.
\end{align}
Its series representation is
\begin{align}
\label{eq:inc_gamma_ser}
 \Gamma(a,z)=\Gamma(a)-z^a \sum_{n=0}^{\infty}\frac{(-z)^{n}}{n!(a+n)}.
\end{align}
The final result is
\begin{align}\label{eq:sol_diff_eq}
  I(\Lambda^2,\gamma)-I(x_1,\gamma) &= \LQ(-\omega)^{\gamma}\left( \Gamma\left(1+\gamma, -t_{\Lambda}\right) - \Gamma\left(1+\gamma,-t_{1}\right)\right)\\
 &=\LQ\,\omega^{\gamma}\sum_{n=0}^{\infty}\frac{(t_{\Lambda})^{1+\gamma+n}-(t_{1})^{1+\gamma+n}}{n!(1+\gamma+n)},
\end{align}
where $t_1=\ln\left(x_1/\LQ\right)$ and $t_\Lambda=\ln\left(\LL/\LQ\right)$.

\section{Alternative UV parametrization}
\label{sec:alt_UV_parametrization}

The parametrization for the perturbative dressing functions is not unique, as the large momentum behavior can be reproduced by several expressions. Here we explore the effects of a different parametrization that shifts the Landau pole to zero \cite{Richardson:1978bt}. This amounts to the following replacement in the dressing functions:
\begin{align}
 G(s)(\omega \ln(y/\Lambda_\mathrm{QCD}^2))^{\delta} \rightarrow G(s)(\omega \ln (1+y/\Lambda_\mathrm{QCD}^2))^{\delta},\\
 Z(s)(\omega \ln(y/\Lambda_\mathrm{QCD}^2))^{\gamma} \rightarrow Z(s)(\omega \ln (1+y/\Lambda_\mathrm{QCD}^2))^{\gamma}.
\end{align}
The calculations of Sec.~\ref{sec:quadDivs} and Appendix~\ref{sec:quad_div_term} can be repeated along the same lines with the variable substitution $t(y)=-\ln(1+y/\LQ)$. The result is
\begin{align}\label{eq:sol_diff_eq_Richardson}
 I^R(&\Lambda^2)-I^R(x_1)=\omega^{\gamma}\int_{x_1}^{\LL} dy \ln(1+y/\Lambda_\mathrm{QCD}^2)^{\gamma}\\
 &=\Lambda_\mathrm{QCD}^2\,\omega^{\gamma}\sum_{n=0}^{\infty}\frac{\ln(1+\LL/\Lambda_\mathrm{QCD}^2)^{1+\gamma+n}-\ln(1+x_1/\Lambda_\mathrm{QCD}^2)^{1+\gamma+n}}{n!(1+\gamma+n)}.
\end{align}
With the Landau pole in the Richardson coupling now at $p^2=0$, we
extend the
integration in the subtraction terms to $x_1=0$ as well. The
contribution from the lower integration bound then vanishes again. The
remaining cutoff dependent parts of both solutions,
eqs.~(\ref{eq:sol_diff_eq})~and~(\ref{eq:sol_diff_eq_Richardson}),
are equivalent for $\Lambda \gg \Lambda_\mathrm{QCD}$ in the UV.
Consequently, the subtraction terms are the same for both
parametrizations, which should be expected as the 
details at lower momenta must be insignificant for a perturbative 
subtraction of spurious divergences.

\bibliographystyle{utphys_mod}
\bibliography{literature_quad_divs}

\providecommand{\href}[2]{#2}\begingroup\raggedright\begin{thebibliography}{10}

\bibitem{vonSmekal:1997is}
L.~von Smekal, R.~Alkofer, and A.~Hauck,
  \href{http://dx.doi.org/10.1103/PhysRevLett.79.3591}{{\em Phys. Rev. Lett.}
  {\bfseries 79} (1997) 3591--3594},
\href{http://arxiv.org/abs/hep-ph/9705242}{{\ttfamily arXiv:hep-ph/9705242}}.

\bibitem{vonSmekal:1997vx}
L.~von Smekal, A.~Hauck, and R.~Alkofer,
  \href{http://dx.doi.org/10.1006/aphy.1998.5806}{{\em Ann. Phys.} {\bfseries
  267} (1998) 1},
\href{http://arxiv.org/abs/hep-ph/9707327}{{\ttfamily arXiv:hep-ph/9707327}}.

\bibitem{Atkinson:1997tu}
D.~Atkinson and J.~C.~R. Bloch,
  \href{http://dx.doi.org/10.1103/PhysRevD.58.094036}{{\em Phys. Rev.}
  {\bfseries D58} (1998) 094036},
\href{http://arxiv.org/abs/hep-ph/9712459}{{\ttfamily arXiv:hep-ph/9712459}}.

\bibitem{Alkofer:2000wg}
R.~Alkofer and L.~von Smekal,
  \href{http://dx.doi.org/10.1016/S0370-1573(01)00010-2}{{\em Phys.Rept.}
  {\bfseries 353} (2001) 281},
\href{http://arxiv.org/abs/hep-ph/0007355}{{\ttfamily arXiv:hep-ph/0007355
  [hep-ph]}}.

\bibitem{Fischer:2002eq}
C.~S. Fischer, R.~Alkofer, and H.~Reinhardt,
  \href{http://dx.doi.org/10.1103/PhysRevD.65.094008}{{\em Phys. Rev.}
  {\bfseries D65} (2002) 094008},
\href{http://arxiv.org/abs/hep-ph/0202195}{{\ttfamily arXiv:hep-ph/0202195}}.

\bibitem{Fischer:2002hn}
C.~S. Fischer and R.~Alkofer,
  \href{http://dx.doi.org/10.1016/S0370-2693(02)01809-9}{{\em Phys. Lett.}
  {\bfseries B536} (2002) 177--184},
\href{http://arxiv.org/abs/hep-ph/0202202}{{\ttfamily arXiv:hep-ph/0202202}}.

\bibitem{Fischer:2003rp}
C.~S. Fischer and R.~Alkofer,
  \href{http://dx.doi.org/10.1103/PhysRevD.67.094020}{{\em Phys. Rev.}
  {\bfseries D67} (2003) 094020},
\href{http://arxiv.org/abs/hep-ph/0301094}{{\ttfamily arXiv:hep-ph/0301094}}.

\bibitem{Fischer:2006ub}
C.~S. Fischer, \href{http://dx.doi.org/10.1088/0954-3899/32/8/R02}{{\em J.
  Phys.} {\bfseries G32} (2006) R253--R291},
\href{http://arxiv.org/abs/hep-ph/0605173}{{\ttfamily arXiv:hep-ph/0605173}}.

\bibitem{Alkofer:2008tt}
R.~Alkofer, C.~S. Fischer, F.~J. Llanes-Estrada, and K.~Schwenzer,
  \href{http://dx.doi.org/10.1016/j.aop.2008.07.001}{{\em Annals Phys.}
  {\bfseries 324} (2009) 106--172},
\href{http://arxiv.org/abs/0804.3042}{{\ttfamily arXiv:0804.3042 [hep-ph]}}.

\bibitem{Boucaud:2008ji}
P.~Boucaud {\em et~al.},
  \href{http://dx.doi.org/10.1088/1126-6708/2008/06/012}{{\em JHEP} {\bfseries
  06} (2008) 012},
\href{http://arxiv.org/abs/0801.2721}{{\ttfamily arXiv:0801.2721 [hep-ph]}}.

\bibitem{Aguilar:2008xm}
A.~Aguilar, D.~Binosi, and J.~Papavassiliou,
  \href{http://dx.doi.org/10.1103/PhysRevD.78.025010}{{\em Phys.Rev.}
  {\bfseries D78} (2008) 025010},
  \href{http://arxiv.org/abs/0802.1870}{{\ttfamily arXiv:0802.1870 [hep-ph]}}.

\bibitem{Fischer:2008uz}
C.~S. Fischer, A.~Maas, and J.~M. Pawlowski,
  \href{http://dx.doi.org/10.1016/j.aop.2009.07.009}{{\em Annals Phys.}
  {\bfseries 324} (2009) 2408--2437},
\href{http://arxiv.org/abs/0810.1987}{{\ttfamily arXiv:0810.1987 [hep-ph]}}.

\bibitem{Boucaud:2010gr}
P.~Boucaud, M.~Gomez, J.~Leroy, A.~Le~Yaouanc, J.~Micheli, {\em et~al.},
  \href{http://dx.doi.org/10.1103/PhysRevD.82.054007}{{\em Phys.Rev.}
  {\bfseries D82} (2010) 054007},
\href{http://arxiv.org/abs/1004.4135}{{\ttfamily arXiv:1004.4135 [hep-ph]}}.

\bibitem{Pennington:2011xs}
M.~Pennington and D.~Wilson,
  \href{http://dx.doi.org/10.1103/PhysRevD.84.094028}{{\em Phys.Rev.}
  {\bfseries D84} (2011) 119901},
\href{http://arxiv.org/abs/1109.2117}{{\ttfamily arXiv:1109.2117 [hep-ph]}}.

\bibitem{Maas:2011se}
A.~Maas, \href{http://dx.doi.org/10.1016/j.physrep.2012.11.002}{{\em
  Phys.Rept.} {\bfseries 524} (2013) 203--300},
\href{http://arxiv.org/abs/1106.3942}{{\ttfamily arXiv:1106.3942 [hep-ph]}}.

\bibitem{Binosi:2012sj}
D.~Binosi, D.~Ib\'a\~nez, and J.~Papavassiliou,
  \href{http://dx.doi.org/10.1103/PhysRevD.86.085033}{{\em Phys.Rev.}
  {\bfseries D86} (2012) 085033},
\href{http://arxiv.org/abs/1208.1451}{{\ttfamily arXiv:1208.1451 [hep-ph]}}.

\bibitem{Strauss:2012dg}
S.~Strauss, C.~S. Fischer, and C.~Kellermann,
  \href{http://dx.doi.org/10.1103/PhysRevLett.109.252001}{{\em Phys.Rev.Lett.}
  {\bfseries 109} (2012) 252001},
\href{http://arxiv.org/abs/1208.6239}{{\ttfamily arXiv:1208.6239 [hep-ph]}}.

\bibitem{Huber:2012kd}
M.~Q. Huber and L.~von Smekal,
  \href{http://dx.doi.org/10.1007/JHEP04(2013)149}{{\em JHEP} {\bfseries 1304}
  (2013) 149},
\href{http://arxiv.org/abs/1211.6092}{{\ttfamily arXiv:1211.6092 [hep-th]}}.

\bibitem{Fischer:2012vc}
C.~S. Fischer and J.~Luecker,
  \href{http://dx.doi.org/10.1016/j.physletb.2012.11.054}{{\em Phys.Lett.}
  {\bfseries B718} (2013) 1036--1043},
\href{http://arxiv.org/abs/1206.5191}{{\ttfamily arXiv:1206.5191 [hep-ph]}}.

\bibitem{Muller:2013pya}
D.~M\"uller, M.~Buballa, and J.~Wambach,
  \href{http://dx.doi.org/10.1140/epja/i2013-13096-5}{{\em Eur.Phys.J.}
  {\bfseries A49} (2013) 96},
\href{http://arxiv.org/abs/1303.2693}{{\ttfamily arXiv:1303.2693 [hep-ph]}}.

\bibitem{Hopfer:2013np}
M.~Hopfer, A.~Windisch, and R.~Alkofer, {\em PoS} {\bfseries ConfinementX}
  (2012) 073,
\href{http://arxiv.org/abs/1301.3672}{{\ttfamily arXiv:1301.3672 [hep-ph]}}.

\bibitem{Muller:2013tya}
D.~M\"uller, M.~Buballa, and J.~Wambach,
  \href{http://dx.doi.org/10.1016/j.physletb.2013.10.050}{{\em Phys.Lett.}
  {\bfseries B727} (2013) 240--243},
\href{http://arxiv.org/abs/1308.4303}{{\ttfamily arXiv:1308.4303 [hep-ph]}}.

\bibitem{Fischer:2013eca}
C.~S. Fischer, L.~Fister, J.~Luecker, and J.~M. Pawlowski,
\href{http://arxiv.org/abs/1306.6022}{{\ttfamily arXiv:1306.6022 [hep-ph]}}.

\bibitem{Aguilar:2013xqa}
A.~Aguilar, D.~Ib\'a\~nez, and J.~Papavassiliou,
  \href{http://dx.doi.org/10.1103/PhysRevD.87.114020}{{\em Phys.Rev.}
  {\bfseries D87} (2013) 114020},
\href{http://arxiv.org/abs/1303.3609}{{\ttfamily arXiv:1303.3609 [hep-ph]}}.

\bibitem{Blum:2014gna}
A.~Blum, M.~Q. Huber, M.~Mitter, and L.~von Smekal,
  \href{http://dx.doi.org/10.1103/PhysRevD.89.061703}{{\em Phys.Rev.}
  {\bfseries D89} (2014) 061703(R)},
\href{http://arxiv.org/abs/1401.0713}{{\ttfamily arXiv:1401.0713 [hep-ph]}}.

\bibitem{Aguilar:2014tka}
A.~Aguilar, D.~Binosi, and J.~Papavassiliou,
\href{http://arxiv.org/abs/1401.3631}{{\ttfamily arXiv:1401.3631 [hep-ph]}}.

\bibitem{Eichmann:2014xya}
G.~Eichmann, R.~Williams, R.~Alkofer, and M.~Vujinovic,
\href{http://arxiv.org/abs/1402.1365}{{\ttfamily arXiv:1402.1365 [hep-ph]}}.

\bibitem{Pascual:1984}
P.~Pascual and R.~Tarrach, {\em QCD: Renormalization for the practitioner},
  vol.~194 of {\em Lecture Notes in Physics}.
\newblock Springer, Heidelberg, 1984.

\bibitem{Gusynin:1998se}
V.~Gusynin, A.~Schreiber, T.~Sizer, and A.~G. Williams,
  \href{http://dx.doi.org/10.1103/PhysRevD.60.065007}{{\em Phys.Rev.}
  {\bfseries D60} (1999) 065007},
\href{http://arxiv.org/abs/hep-th/9811184}{{\ttfamily arXiv:hep-th/9811184
  [hep-th]}}.

\bibitem{Phillips:1999bf}
D.~R. Phillips, I.~Afnan, and A.~Henry-Edwards,
  \href{http://dx.doi.org/10.1103/PhysRevC.61.044002}{{\em Phys.Rev.}
  {\bfseries C61} (2000) 044002},
\href{http://arxiv.org/abs/nucl-th/9910063}{{\ttfamily arXiv:nucl-th/9910063
  [nucl-th]}}.

\bibitem{Dudal:2007cw}
D.~Dudal, S.~P. Sorella, N.~Vandersickel, and H.~Verschelde,
  \href{http://dx.doi.org/10.1103/PhysRevD.77.071501}{{\em Phys. Rev.}
  {\bfseries D77} (2008) 071501},
\href{http://arxiv.org/abs/0711.4496}{{\ttfamily arXiv:0711.4496 [hep-th]}}.

\bibitem{Alkofer:2008jy}
R.~Alkofer, M.~Q. Huber, and K.~Schwenzer,
  \href{http://dx.doi.org/http://link.aps.org/doi/10.1103/PhysRevD.81.105010}{{\em
  Phys. Rev.} {\bfseries D81} (2010) 105010},
\href{http://arxiv.org/abs/0801.2762}{{\ttfamily arXiv:0801.2762 [hep-th]}}.

\bibitem{Pawlowski:2003hq}
J.~M. Pawlowski, D.~F. Litim, S.~Nedelko, and L.~von Smekal,
  \href{http://dx.doi.org/10.1103/PhysRevLett.93.152002}{{\em Phys. Rev. Lett.}
  {\bfseries 93} (2004) 152002},
\href{http://arxiv.org/abs/hep-th/0312324}{{\ttfamily arXiv:hep-th/0312324}}.

\bibitem{Lerche:2002ep}
C.~Lerche and L.~von Smekal,
  \href{http://dx.doi.org/10.1103/PhysRevD.65.125006}{{\em Phys.Rev.}
  {\bfseries D65} (2002) 125006},
\href{http://arxiv.org/abs/hep-ph/0202194}{{\ttfamily arXiv:hep-ph/0202194}}.

\bibitem{Huber:2012zj}
M.~Q. Huber, A.~Maas, and L.~von Smekal,
  \href{http://dx.doi.org/10.1007/JHEP11(2012)035}{{\em JHEP} {\bfseries 1211}
  (2012) 035},
\href{http://arxiv.org/abs/1207.0222}{{\ttfamily arXiv:1207.0222 [hep-th]}}.

\bibitem{Zwanziger:2001kw}
D.~Zwanziger, {\em Phys. Rev.} {\bfseries D65} (2002) 094039,
\href{http://arxiv.org/abs/hep-th/0109224}{{\ttfamily arXiv:hep-th/0109224}}.

\bibitem{Cucchieri:2007md}
A.~Cucchieri and T.~Mendes, {\em PoS} {\bfseries LAT2007} (2007) 297,
\href{http://arxiv.org/abs/0710.0412}{{\ttfamily arXiv:0710.0412 [hep-lat]}}.

\bibitem{Cucchieri:2008fc}
A.~Cucchieri and T.~Mendes,
  \href{http://dx.doi.org/10.1103/PhysRevD.78.094503}{{\em Phys. Rev.}
  {\bfseries D78} (2008) 094503},
\href{http://arxiv.org/abs/0804.2371}{{\ttfamily arXiv:0804.2371 [hep-lat]}}.

\bibitem{Sternbeck:2007ug}
A.~Sternbeck, L.~von Smekal, D.~Leinweber, and A.~Williams, {\em PoS}
  {\bfseries LAT2007} (2007) 340,
\href{http://arxiv.org/abs/0710.1982}{{\ttfamily arXiv:0710.1982 [hep-lat]}}.

\bibitem{Bogolubsky:2009dc}
I.~L. Bogolubsky, E.~M. Ilgenfritz, M.~M\"uller-Preussker, and A.~Sternbeck,
  \href{http://dx.doi.org/10.1016/j.physletb.2009.04.076}{{\em Phys. Lett.}
  {\bfseries B676} (2009) 69--73},
\href{http://arxiv.org/abs/0901.0736}{{\ttfamily arXiv:0901.0736 [hep-lat]}}.

\bibitem{Oliveira:2012eh}
O.~Oliveira and P.~J. Silva,
  \href{http://dx.doi.org/10.1103/PhysRevD.86.114513}{{\em Phys.Rev.}
  {\bfseries D86} (2012) 114513},
\href{http://arxiv.org/abs/1207.3029}{{\ttfamily arXiv:1207.3029 [hep-lat]}}.

\bibitem{Sternbeck:2012mf}
A.~Sternbeck and M.~M\"ller-Preussker,
  \href{http://dx.doi.org/10.1016/j.physletb.2013.08.017}{{\em Phys.Lett.}
  {\bfseries B726} (2013) 396--403},
\href{http://arxiv.org/abs/1211.3057}{{\ttfamily arXiv:1211.3057 [hep-lat]}}.

\bibitem{vonSmekal:2008ws}
L.~von Smekal, \href{http://arxiv.org/abs/0812.0654}{{\ttfamily arXiv:0812.0654
  [hep-th]}}.
Plenary talk at 13th International Conference on Selected Problems of Modern
  Theoretical Physics (SPMTP 08), Dubna, Russia, 23-27 Jun 2008.,

\bibitem{Fischer:2010is}
C.~S. Fischer and L.~von Smekal,
  \href{http://dx.doi.org/10.1063/1.3574991}{{\em AIP Conf.Proc.} {\bfseries
  1343} (2011) 247--249},
\href{http://arxiv.org/abs/1011.6482}{{\ttfamily arXiv:1011.6482 [hep-ph]}}.

\bibitem{Curci:1976ar}
G.~Curci and R.~Ferrari,
\href{http://dx.doi.org/10.1016/0370-2693(76)90475-5}{{\em Phys.Lett.}
  {\bfseries B63} (1976) 91--94}.

\bibitem{Baulieu:1981sb}
L.~Baulieu and J.~Thierry-Mieg,
\href{http://dx.doi.org/10.1016/0550-3213(82)90454-0}{{\em Nucl. Phys.}
  {\bfseries B197} (1982) 477}.

\bibitem{ThierryMieg:1985yv}
J.~Thierry-Mieg,
\href{http://dx.doi.org/10.1016/0550-3213(85)90562-0}{{\em Nucl. Phys.}
  {\bfseries B261} (1985) 55}.

\bibitem{Alkofer:2003jr}
R.~Alkofer, C.~S. Fischer, H.~Reinhardt, and L.~von Smekal,
  \href{http://dx.doi.org/10.1103/PhysRevD.68.045003}{{\em Phys. Rev.}
  {\bfseries D68} (2003) 045003},
\href{http://arxiv.org/abs/hep-th/0304134}{{\ttfamily arXiv:hep-th/0304134}}.

\bibitem{vonSmekal:2008en}
L.~von Smekal, M.~Ghiotti, and A.~G. Williams,
  \href{http://dx.doi.org/10.1103/PhysRevD.78.085016}{{\em Phys.Rev.}
  {\bfseries D78} (2008) 085016},
\href{http://arxiv.org/abs/0807.0480}{{\ttfamily arXiv:0807.0480 [hep-th]}}.

\bibitem{vonSmekal:2009ae}
L.~von Smekal, K.~Maltman, and A.~Sternbeck,
  \href{http://dx.doi.org/10.1016/j.physletb.2009.10.030}{{\em Phys.Lett.}
  {\bfseries B681} (2009) 336--342},
\href{http://arxiv.org/abs/0903.1696}{{\ttfamily arXiv:0903.1696 [hep-ph]}}.

\bibitem{Wschebor:2007vh}
N.~Wschebor, \href{http://dx.doi.org/10.1142/S0217751X08040469}{{\em
  Int.J.Mod.Phys.} {\bfseries A23} (2008) 2961--2973},
\href{http://arxiv.org/abs/hep-th/0701127}{{\ttfamily arXiv:hep-th/0701127
  [hep-th]}}.

\bibitem{Ilgenfritz:2006he}
E.~M. Ilgenfritz, M.~M{\"u}ller-Preussker, A.~Sternbeck, A.~Schiller, and I.~L.
  Bogolubsky, {\em Braz. J. Phys.} {\bfseries 37} (2007) 193,
\href{http://arxiv.org/abs/hep-lat/0609043}{{\ttfamily arXiv:hep-lat/0609043}}.

\bibitem{Cucchieri:2008qm}
A.~Cucchieri, A.~Maas, and T.~Mendes,
  \href{http://dx.doi.org/10.1103/PhysRevD.77.094510}{{\em Phys. Rev.}
  {\bfseries D77} (2008) 094510},
\href{http://arxiv.org/abs/0803.1798}{{\ttfamily arXiv:0803.1798 [hep-lat]}}.

\bibitem{Schleifenbaum:2004id}
W.~Schleifenbaum, A.~Maas, J.~Wambach, and R.~Alkofer,
  \href{http://dx.doi.org/10.1103/PhysRevD.72.014017}{{\em Phys.Rev.}
  {\bfseries D72} (2005) 014017},
\href{http://arxiv.org/abs/hep-ph/0411052}{{\ttfamily arXiv:hep-ph/0411052
  [hep-ph]}}.

\bibitem{Boucaud:2011eh}
P.~Boucaud, D.~Dudal, J.~Leroy, O.~Pene, and J.~Rodriguez-Quintero,
  \href{http://dx.doi.org/10.1007/JHEP12(2011)018}{{\em JHEP} {\bfseries 1112}
  (2011) 018},
\href{http://arxiv.org/abs/1109.3803}{{\ttfamily arXiv:1109.3803 [hep-ph]}}.

\bibitem{Fister:2011uw}
L.~Fister and J.~M. Pawlowski,
\href{http://arxiv.org/abs/1112.5440}{{\ttfamily arXiv:1112.5440 [hep-ph]}}.

\bibitem{Pelaez:2013cpa}
M.~Pelaez, M.~Tissier, and N.~Wschebor,
  \href{http://dx.doi.org/10.1103/PhysRevD.88.125003}{{\em Phys.Rev.}
  {\bfseries D88} (2013) 125003},
\href{http://arxiv.org/abs/1310.2594}{{\ttfamily arXiv:1310.2594 [hep-th]}}.

\bibitem{Brown:1988bm}
N.~Brown and M.~Pennington,
\href{http://dx.doi.org/10.1103/PhysRevD.38.2266}{{\em Phys.Rev.} {\bfseries
  D38} (1988) 2266}.

\bibitem{Brown:1988bn}
N.~Brown and M.~Pennington,
\href{http://dx.doi.org/10.1103/PhysRevD.39.2723}{{\em Phys.Rev.} {\bfseries
  D39} (1989) 2723}.

\bibitem{Maas:2005hs}
A.~Maas, J.~Wambach, and R.~Alkofer,
  \href{http://dx.doi.org/10.1140/epjc/s2005-02279-8}{{\em Eur. Phys. J.}
  {\bfseries C42} (2005) 93--107},
\href{http://arxiv.org/abs/hep-ph/0504019}{{\ttfamily arXiv:hep-ph/0504019}}.

\bibitem{Cucchieri:2007ta}
A.~Cucchieri, A.~Maas, and T.~Mendes,
  \href{http://dx.doi.org/10.1103/PhysRevD.75.076003}{{\em Phys. Rev.}
  {\bfseries D75} (2007) 076003},
\href{http://arxiv.org/abs/hep-lat/0702022}{{\ttfamily arXiv:hep-lat/0702022}}.

\bibitem{Fischer:2005en}
C.~Fischer, P.~Watson, and W.~Cassing,
  \href{http://dx.doi.org/10.1103/PhysRevD.72.094025}{{\em Phys.Rev.}
  {\bfseries D72} (2005) 094025},
\href{http://arxiv.org/abs/hep-ph/0509213}{{\ttfamily arXiv:hep-ph/0509213
  [hep-ph]}}.

\bibitem{Meyers:2014iwa}
J.~Meyers and E.~S. Swanson,
\href{http://arxiv.org/abs/1403.4350}{{\ttfamily arXiv:1403.4350 [hep-ph]}}.

\bibitem{Schreiber:1998ht}
A.~W. Schreiber, T.~Sizer, and A.~G. Williams,
  \href{http://dx.doi.org/10.1103/PhysRevD.58.125014}{{\em Phys.Rev.}
  {\bfseries D58} (1998) 125014},
\href{http://arxiv.org/abs/hep-ph/9804385}{{\ttfamily arXiv:hep-ph/9804385
  [hep-ph]}}.

\bibitem{Aguilar:2009ke}
A.~C. Aguilar and J.~Papavassiliou,
  \href{http://dx.doi.org/10.1103/PhysRevD.81.034003}{{\em Phys.Rev.}
  {\bfseries D81} (2010) 034003},
\href{http://arxiv.org/abs/0910.4142}{{\ttfamily arXiv:0910.4142 [hep-ph]}}.

\bibitem{Binosi:2009qm}
D.~Binosi and J.~Papavassiliou,
  \href{http://dx.doi.org/10.1016/j.physrep.2009.05.001}{{\em Phys. Rept.}
  {\bfseries 479} (2009) 1--152},
\href{http://arxiv.org/abs/0909.2536}{{\ttfamily arXiv:0909.2536 [hep-ph]}}.

\bibitem{Aguilar:2013vaa}
A.~Aguilar, D.~Binosi, D.~Ib\'a\~nez, and J.~Papavassiliou,
\href{http://arxiv.org/abs/1312.1212}{{\ttfamily arXiv:1312.1212 [hep-ph]}}.

\bibitem{Huber:2011qr}
M.~Q. Huber and J.~Braun,
  \href{http://dx.doi.org/10.1016/j.cpc.2012.01.014}{{\em Comput.Phys.Commun.}
  {\bfseries 183} (2012) 1290--1320},
\href{http://arxiv.org/abs/1102.5307}{{\ttfamily arXiv:1102.5307 [hep-th]}}.

\bibitem{Alkofer:2008nt}
R.~Alkofer, M.~Q. Huber, and K.~Schwenzer,
  \href{http://dx.doi.org/10.1016/j.cpc.2008.12.009}{{\em Comput. Phys.
  Commun.} {\bfseries 180} (2009) 965--976},
\href{http://arxiv.org/abs/0808.2939}{{\ttfamily arXiv:0808.2939 [hep-th]}}.

\bibitem{Huber:2011xc}
M.~Q. Huber and M.~Mitter,
  \href{http://dx.doi.org/10.1016/j.cpc.2012.05.019}{{\em Comput.Phys.Commun.}
  {\bfseries 183} (2012) 2441--2457},
\href{http://arxiv.org/abs/1112.5622}{{\ttfamily arXiv:1112.5622 [hep-th]}}.

\bibitem{Bloch:2003yu}
J.~C. Bloch, \href{http://dx.doi.org/10.1007/s00601-003-0013-3}{{\em Few Body
  Syst.} {\bfseries 33} (2003) 111--152},
\href{http://arxiv.org/abs/hep-ph/0303125}{{\ttfamily arXiv:hep-ph/0303125}}.

\bibitem{Mader:2013ru}
V.~Mader and R.~Alkofer, {\em PoS} {\bfseries ConfinementX} (2012) 063,
\href{http://arxiv.org/abs/1301.7498}{{\ttfamily arXiv:1301.7498 [hep-th]}}.

\bibitem{Sternbeck:2006rd}
A.~Sternbeck, \href{http://arxiv.org/abs/hep-lat/0609016}{{\ttfamily
  arXiv:hep-lat/0609016}}, PhD thesis, Humboldt-Universit\"at zu Berlin,
2006.

\bibitem{Richardson:1978bt}
J.~L. Richardson,
\href{http://dx.doi.org/10.1016/0370-2693(79)90753-6}{{\em Phys.Lett.}
  {\bfseries B82} (1979) 272}.

\end{thebibliography}\endgroup

\end{document}